\documentclass[sigconf,screen,review=false]{acmart}
\renewcommand\footnotetextcopyrightpermission[1]{} 
\usepackage[utf8]{inputenc}
\usepackage[T1]{fontenc}
\usepackage{microtype}
\usepackage{balance}

\usepackage{amsmath}
\numberwithin{equation}{section}
\usepackage{braket}
\usepackage{stfloats}

\renewcommand\footnotetextcopyrightpermission[1]{} 
\usepackage{subcaption}

\usepackage{hyphenat}
\usepackage[ruled,vlined,linesnumbered]{algorithm2e}
\SetArgSty{textnormal}
\SetVlineSkip{0.1em}
\makeatletter
\newcommand{\removelatexerror}{\let\@latex@error\@gobble}
\newcommand{\nosemic}{\renewcommand{\@endalgocfline}{\relax}}
\newcommand{\dosemic}{\renewcommand{\@endalgocfline}{\algocf@endline}}
\let\oldnl\nl
\newcommand{\nonl}{\renewcommand{\nl}{\let\nl\oldnl}}

\makeatother

\SetCommentSty{mycomment}

\usepackage{color}
\usepackage{pifont}
\def\C#1{\raisebox{-.4pt}{\ding{\numexpr171+#1\relax}}}

\def\sC#1{\raisebox{-.4pt}{\ding{\numexpr191+#1\relax}}}
\def\sBC#1{\raisebox{-.4pt}{\ding{\numexpr201+#1\relax}}}

\usepackage{amsthm}
\newtheoremstyle{mytheorem}
    {0.2em}                
    {0.2em}                
    {\slshape}        
    {}                
    {\hspace*{\parindent}\bfseries}       
    {}               
    { }               
    {\thmname{#1}\thmnumber{ #2}\thmnote{ (#3)}}                
    \theoremstyle{mytheorem}
    \newtheorem{theorem}{Theorem}[section]

\newtheoremstyle{mycorollary}
    {0.2em}                
    {0.2em}                
    {\slshape}        
    {}                
    {\hspace*{\parindent}\bfseries}       
    {}               
    { }               
    {\thmname{#1}\thmnumber{ #2}\thmnote{ (#3)}}                
    \theoremstyle{mycorollary}

\newtheoremstyle{mydefinition}
    {0.2em minus 0.2em}                
    {0.2em}                
    {\upshape}        
    {}                
    {\hspace*{\parindent}\bfseries}       
    {}               
    { }               
    {\thmname{#1}\thmnumber{ #2}\thmnote{ (#3)}}                
    \theoremstyle{mydefinition}
    \newtheorem{definition}{Definition}[section]

\newtheoremstyle{mynote}
    {0.2em}                
    {0.2em}                
    {\upshape}        
    {}                
    {\hspace*{\parindent}\bfseries}       
    {}               
    { }               
    {\thmnote{#3}}                
    \theoremstyle{mynote}
    \newtheorem{note}{Note}

\newtheoremstyle{myproof}
    {0.2em}                
    {0.2em}                
    {\upshape}        
    {}                
    {\itshape}       
    {}               
    { }               
    {Proof.}                
    \theoremstyle{myproof}
    \newtheorem{myproof}{Proof}

\newcommand{\myparagraph}[1]{\noindent{\normalfont\normalsize\bfseries #1.}}
\newcommand{\myfirstparagraph}[1]{\noindent{\normalfont\normalsize\bfseries #1.}}
\newcommand{\myleading}[1]{{\normalfont\normalsize\bfseries #1:}}

\newcommand\nb[1]{{\scriptscriptstyle (#1)}}
\newcommand\lb[1]{{{\scriptscriptstyle (} #1 {\scriptscriptstyle )}}}

\newcommand{\CNOT}{\ensuremath{\text{CNOT}}}

\AtBeginDocument{%
    \setlength{\abovedisplayskip}{4pt}
    \setlength{\belowdisplayskip}{5pt}
    \setlength{\abovedisplayshortskip}{0pt}
    \setlength{\belowdisplayshortskip}{5pt}
}

\begin{document}

\title{Suppressing ZZ Crosstalk of Quantum Computers through Pulse and Scheduling Co-Optimization}

\author{Lei Xie}
\affiliation{%
  \institution{Tsinghua University}
  \city{Beijing}
  \country{China}}
\email{xie-l18@mails.tsinghua.edu.cn}

\author{Jidong Zhai}
\affiliation{%
  \institution{Tsinghua University}
  \city{Beijing}
  \country{China}}
\email{zhaijidong@tsinghua.edu.cn}

\author{ZhenXing Zhang}
\affiliation{%
  \institution{Tencent Quantum Laboratory}
  \city{Shenzhen}
  \country{China}}
\email{alanzxzhang@tencent.com}

\author{Jonathan Allcock}
\affiliation{%
  \institution{Tencent Quantum Laboratory}
  \city{Shenzhen}
  \country{China}}
\email{jonallcock@tencent.com}

\author{Shengyu Zhang}
\affiliation{%
  \institution{Tencent Quantum Laboratory}
  \city{Shenzhen}
  \country{China}}
\email{shengyzhang@tencent.com}

\author{Yi-Cong Zheng}
\authornote{Corresponding author}
\affiliation{%
  \institution{Tencent Quantum Laboratory}
  \city{Shenzhen}
  \country{China}}
\email{yicongzheng@tencent.com}


\begin{abstract}

Noise is a significant obstacle to quantum computing, and $ZZ$ crosstalk is one of the most destructive types of noise affecting superconducting qubits. Previous approaches to suppressing $ZZ$ crosstalk have mainly relied on specific chip design that can complicate chip fabrication and aggravate decoherence. To some extent, special chip design can be avoided by relying on pulse optimization to suppress $ZZ$ crosstalk. However, existing approaches are non-scalable, as their required time and memory grow exponentially with the number of qubits involved.

To address the above problems, we propose a scalable approach by co-optimizing pulses and scheduling. We optimize pulses to offer an ability to suppress $ZZ$ crosstalk surrounding a gate, and then design scheduling strategies to exploit this ability and achieve suppression across the whole circuit. A main advantage of such co-optimization is that it does not require special hardware support. Besides, we implement our approach as a general framework that is compatible with different pulse optimization methods. We have conducted extensive evaluations by simulation and on a real quantum computer. Simulation results show that our proposal can improve the fidelity of quantum computing on $4{\sim}12$ qubits by up to $81\times$ ($11\times$ on average). Ramsey experiments on a real quantum computer also demonstrate that our method can eliminate the effect of $ZZ$ crosstalk to a great extent.

\end{abstract}


\maketitle
\pagestyle{plain}

\section{Introduction}
\label{sec:introduction}

Quantum computing (QC), in theory, can be used to solve some classically difficult or intractable problems~\cite{shor1999polynomial,arute2019quantum}. However, current quantum computers are susceptible to noise, which impedes their usage in many scenarios~\cite{murali2019noise,murali2019full}. Noise can corrupt qubits, degrade quantum gate fidelity, and lead to computational errors. Finding effective approaches to suppressing noise is therefore crucial for enabling near-term quantum computers to solve practical problems.

\begin{figure}[!h]
    \centering
    \includegraphics[width=0.45\columnwidth]{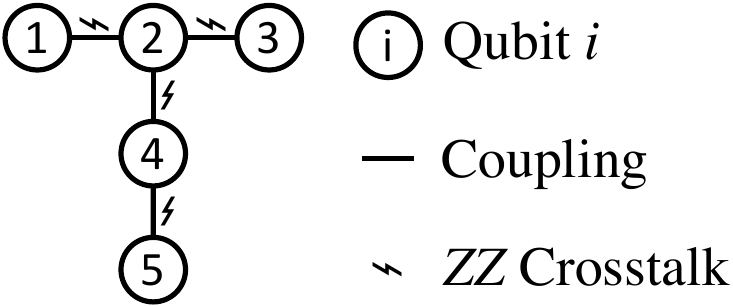}
    \caption{IBMQ Vigo superconducting device topology. Couplings between qubits mediate $ZZ$ crosstalk.}
    \label{fig:vigo}
\end{figure}

Superconducting qubits are one of the leading technologies for building quantum computers. However, devices based on this technology are affected by a destructive type of noise known as $ZZ$ crosstalk~\cite{diCarlo2009demonstration,mckay2019three,cai2021impact}. This refers to an always-on $\sigma_z \otimes \sigma_z$ interaction between \textit{qubits connected by couplings} (Figure~\ref{fig:vigo}), which originates from the interaction between the computational and non-computational energy levels of qubits~\cite{diCarlo2009demonstration,arute2019quantum}.

$ZZ$ crosstalk is problematic for QC in several ways. Firstly, its presence is often a limiting factor in quantum gate fidelity~\cite{mundada2019suppression}, leading to a ${\sim}5\times$ increase in terms of the error rate of two-qubit gates~\cite{ku2020suppression,sung2020realization,han2020error}. Secondly, it is also an obstacle limiting the parallel execution of gates in a quantum circuit: when multiple gates are performed simultaneously, $ZZ$ crosstalk can further deteriorate the fidelity of each gate by $2{\sim}10\times$~\cite{mckay2019three,li2020tunable}. Additionally, $ZZ$ crosstalk can cause correlated high-weight errors to occur with high probability~\cite{sarovar2020detecting,mckay2019three}, which is difficult for quantum error correction (QEC) to deal with~\cite{aliferis2006quantum,fowler2012surface,lidar2013quantum}.

Previous approaches to suppressing $ZZ$ crosstalk have mostly relied on sophisticated and specific chip design. One widely explored approach is to use tunable couplers~\cite{niskanen2007quantum,yan2018tunable,li2020tunable,sung2020realization}, but this introduces additional decoherence factors~\cite{kandala2020demonstration,malekakhlagh2020first}, shortening the lifetime of qubits. Recently, two other approaches based on heterogeneous qubits~\cite{ku2020suppression,zhao2020high,noguchi2020fast} and multiple coupling paths~\cite{mundada2019suppression,kandala2020demonstration} have been proposed. However, these approaches suffer from common drawbacks of the extra complexity imposed on chip fabrication and control, and of the fact that none of them can be used on simpler devices with homogeneous qubits connected by a single untunable coupling (e.g., IBMQ devices~\cite{corcoles2015demonstration,takita2016demonstration}) which are expected to have much longer coherence time~\cite{takita2016demonstration,gambetta2017building}.

Superconducting quantum computers use microwave pulses to control qubits~\cite{arute2019quantum,mckay2017efficient,motzoi2009simple}, and it is thus desirable to suppress $ZZ$ crosstalk via careful manipulation of these pulses, instead of via specific chip design. Since $ZZ$ crosstalk causes error correlation, to minimize its effect on the execution of quantum circuits, an ideal approach is to treat the whole circuit as a single unitary and optimize pulses globally~\cite{khaneja2005optimal,d2007introduction}. However, such kind of global optimization is limited to small circuits, as \emph{its time and memory consumption grows exponentially with the number of qubits involved}~\cite{leung2017speedup,shi2019optimized}.

To overcome the above problems, we propose \emph{co-optimizing pulses and scheduling}. As shown in Figure~\ref{fig:workflow}, we optimize pulses to implement quantum gates and offer an ability to suppress $ZZ$ crosstalk \textit{surrounding the gate}, and then we design effective scheduling strategies for exploiting this ability to achieve suppression \textit{across the whole circuit}.

We manage to tackle $ZZ$ crosstalk in a scalable way. (1) We carefully design the objective of pulse optimization so that we only need to optimize pulses \textit{on small systems} (typically with $\leq 4$ qubits), which has low time and memory consumption. (2) We leverage the duality between cuts and odd-vertex pairings of graphs and several heuristics to design a scheduler that can achieve effective suppression in time \textit{polynomial} in the number of qubits and gates of quantum circuits.

We implement our approach as a general framework that can work with different pulse optimization methods (we have tested two existing methods and one proposed in this work). Simulation results show that we can improve the fidelity of QC with $4{\sim}12$ qubits by up to $81\times$ ($11\times$ on average), compared with the state-of-the-art pulse and scheduling policy used on current quantum computers. Ramsey experiments on a real QC device with three transmon qubits also demonstrate that our method can eliminate the effect of $ZZ$ crosstalk to a great extent by reducing effective $ZZ$ strength (defined in Section~\ref{sec:ramsey}) from ${\sim}200$ kHz to <11 kHz.


\section{Overview and Challenges}
\label{sec:overview}

\begin{figure}[!t]
    \centering
    \includegraphics[width=\columnwidth]{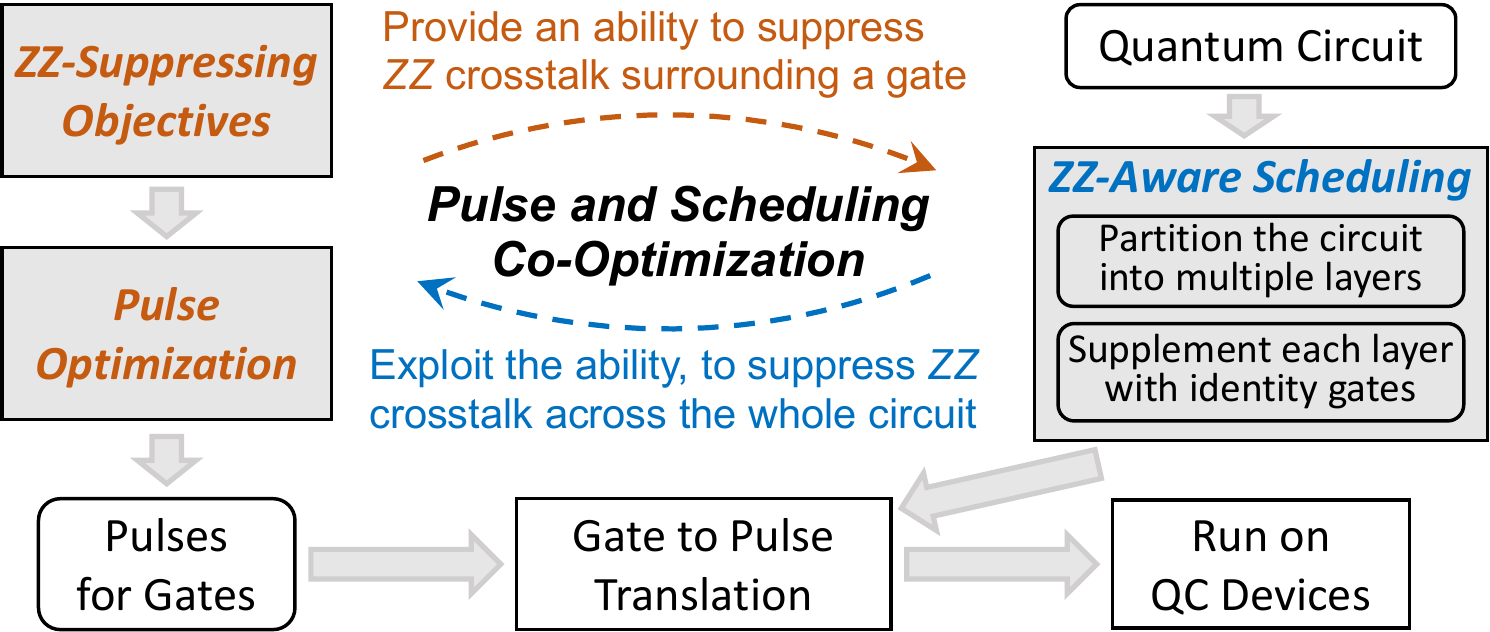}
    \caption{Our pulse and scheduling co-optimization approach.}
    \label{fig:workflow}
\end{figure}

The key idea of our approach is to co-optimize pulses and scheduling so as to suppress $ZZ$ crosstalk in a scalable way. As shown in Figure~\ref{fig:workflow}, given a quantum circuit, our $ZZ$-aware scheduling first partitions it into multiple layers\footnote{A layer contains a set of gates to be executed simultaneously and different layers are executed serially.} and then supplements each layer with additional identity gates. Next, we translate each gate in the circuit to its corresponding pulses that are optimized under our $ZZ$-suppressing objectives. In the following, we use an example to illustrate our idea.

\subsection{A Motivating Example}

\begin{figure*}[!t]
    \centering
    \includegraphics[width=\linewidth]{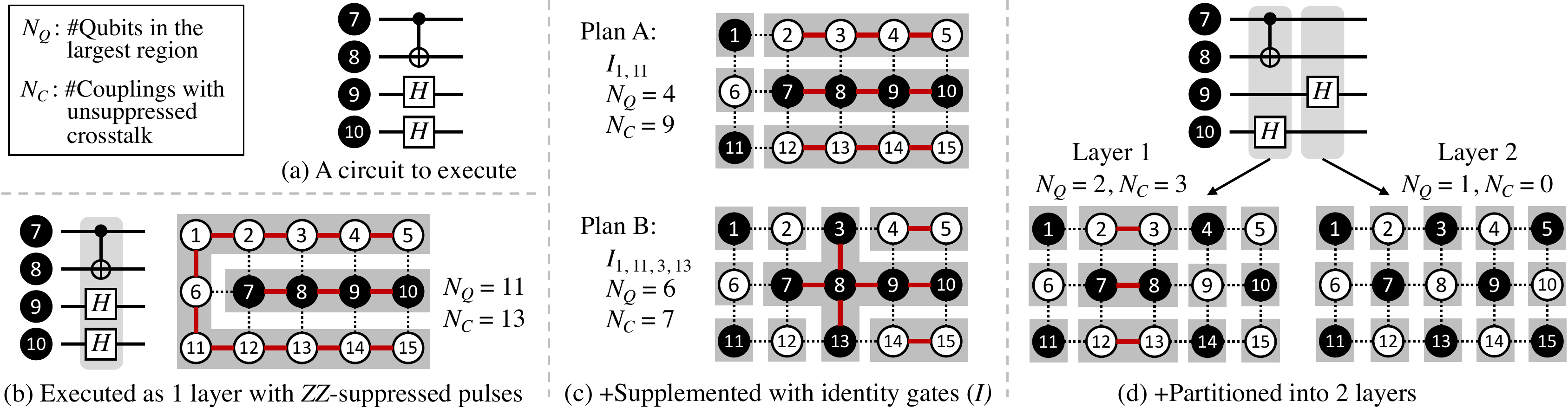}
    \caption{A motivating example. With the introduction of new strategies to (c) and (d), suppression is gradually enhanced (indicated by smaller $N_Q$ and $N_C$). Black qubits have pulses applied to, while white ones do not. Gray areas of topology graphs represent regions. Crosstalk across regions (dashed edges) is suppressed but that within regions (solid edges) is not.}
    \label{fig:overview}
\end{figure*}

Figure~\ref{fig:overview} shows an example where we aim to run a circuit (a) on a device with a $5 \times 3$ grid topology.

\myparagraph{Optimizing pulses for quantum gates} For now, we consider executing all the 3 gates of (a) simultaneously as 1 layer (Figure~\ref{fig:overview}~(b)). According to qubit status (whether a gate is applied to the qubit or not), we divide the device into two \textit{regions} (marked by gray areas), where a region refers to a connected component composed of qubits with the same status. \textit{We optimize pulses to implement quantum gates while also suppressing cross-region $ZZ$ crosstalk surrounding the gates.} For example, the optimized pulses on qubits 7-10 implement gates $\CNOT_{7,8}, H_{9,10}$ and suppress $ZZ$ crosstalk on the 9 cross-region couplings (dashed edges) surrounding them.

However, there are still 13 \textit{intra-region} couplings (solid edges), and these are the ones with unsuppressed $ZZ$ crosstalk. We use two metrics to characterize them: \#couplings with unsuppressed crosstalk ($N_C$) and \#qubits in the largest region ($N_Q$). A larger $N_C$ implies more crosstalk errors, and a larger $N_Q$ can produce correlated errors of higher weight.~\cite{aliferis2006quantum}. We propose two scheduling strategies to reduce these metrics.

\myparagraph{Applying additional identity gates} The first strategy is to \textit{apply identity gates to some of the qubits on which no gates have been performed.} Identity gates (pulses) do not alter the quantum state of qubits, but they can convert some intra-region crosstalk to cross-region crosstalk that can be suppressed by these identity pulses. Figure~\ref{fig:overview}~(c) gives two plans. Take Plan A for example, it applies identity gates on qubits 1 and 11, converting unsuppressed intra-region crosstalk 1-2, 1-6, 11-6 and 11-12 to suppressed cross-region crosstalk.

\myparagraph{Partitioning a circuit into multiple layers} Besides applying identity gates, suppression can be further enhanced if we \textit{properly partition the circuit into 2 layers} instead of one, as shown in Figure~\ref{fig:overview}~(d). Compared with (c), both $N_Q$ and $N_C$ of each layer are largely reduced, which indicates much better suppression. As an intuitive demonstration, all the $ZZ$ crosstalk on the device is suppressed in Layer 2. 

\subsection{Challenges}

We have addressed three main challenges for applying our approach in practice.

\textbf{The first challenge is that pulse optimization requires time and memory exponential in the number of qubits involved.} We address this by setting the objective of pulse optimization to suppressing only cross-region $ZZ$ crosstalk. Section~\ref{sec:gate_lvl} elaborates on the objective and explains why we only need to optimize pulses on small systems (i.e., with only a few qubits) under this objective.

\textbf{The second challenge is to find a plan to apply identity gates with the resulting $N_Q$ and $N_C$ minimized.} It is challenging as the minimization of $N_C$ alone can be reduced to a maximum cut problem (see Section~\ref{sec:layer_lvl}) which is NP-complete for general graphs. Besides, Plan A ($N_Q=4, N_C=9$) and B ($N_Q=6, N_C=7$) in Figure~\ref{fig:overview}~(c) also show a trade-off between $N_Q$ and $N_C$, and thus consideration of $N_Q$ further complicates the situation. In Section~\ref{sec:layer_lvl}, we define an \textit{optimal suppression} problem to characterize this trade-off. And based on the duality of cuts and odd-vertex pairings, we give an \textit{efficient} solution for near-term QC devices with a planar topology.

\textbf{The third challenge is to strike a balance between parallelism and crosstalk suppression when partitioning circuits.} Though, as shown in Figure~\ref{fig:overview}~(d), more layers can improve suppression, they can also decrease parallelism and hence increase the effect of decoherence. For high-fidelity QC, we need to balance these two factors. The challenge is that it is impractical to test all possible partitioning schemes and find an optimal one. We instead propose several heuristics which, according to our evaluations, give very good performance in practice. Section~\ref{sec:circuit_lvl} describes these heuristics along with a complete scheduling algorithm.

\section{Preliminaries}
\label{sec:background}

\subsection{Quantum Evolution \& Quantum Control}
\label{sec:qe}

QC is performed by using quantum gates to manipulate qubits, whose evolution is described by the Schr{\"o}dinger equation
\begin{equation*}
    i \hbar \ket{\dot{\psi}(t)} = H(t) \ket{\psi(t)}
\end{equation*}
where $\ket{\psi(t)}$ is a state of the qubits, $\ket{\dot{\psi}(t)}$ is its derivative with respect to time $t$, and $H(t)$ is the Hamiltonian that completely determines the evolution. We discuss our approach based on the effective Hamiltonian below~\cite{magesan2020effective,leung2017speedup}. Drive noise and leakage into higher energy levels will be considered in the evaluation.

\begin{figure}[!h]
    \centering
    \begin{minipage}[c]{0.2\columnwidth}
        \centering
        \includegraphics[width=0.65\columnwidth]{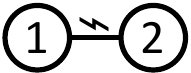}
    \end{minipage}
    ~
    \begin{minipage}[c]{0.7\columnwidth}
        \centering
        \scalebox{0.95}{%
        $\begin{aligned}
            H(t) &= \lambda \sigma_z^\nb{1} \otimes \sigma_z^\nb{2} \\
                &+ \textstyle{\sum}_{q \in \{1,2\}} \left[ \Omega^\lb{q}_x(t) \sigma^\lb{q}_x + \Omega^\lb{q}_y(t) \sigma^\lb{q}_y \right] \\
                &+ \Omega^\nb{1\text{-}2}(t) H_{Coupling}
        \end{aligned}$%
        }
    \end{minipage}
    \caption{A two-qubit system and its (effective) Hamiltonian.}
    \label{fig:evolution}
\end{figure}

In superconducting QC devices, quantum gates are implemented using pulses to control certain Hamiltonian terms~\cite{mckay2017efficient}. For the two-qubit system in Figure~\ref{fig:evolution}, $\sigma^\nb{1}_z \otimes \sigma^\nb{2}_z$ describes $ZZ$ crosstalk, $\lambda$ is its strength, and $\Omega^\lb{q}_{x,y}(t)$ and $\Omega^\nb{1\text{-}2}(t)$ are control pulses. Single-qubit gates on qubit $q$ can be achieved by controlling the $\sigma^\lb{q}_x$ and $\sigma^\lb{q}_y$ terms via $\Omega^\lb{q}_{x,y}(t)$. Two-qubit gates can be achieved by tuning $\Omega^\nb{1\text{-}2}(t)$ for $H_{Coupling}$. The concrete form of $H_{Coupling}$ depends on the specific device and gate type. For example, $\sigma_z \otimes \sigma_x$ is used to implement CNOT gate~\cite{rigetti2010fully,chow2011simple} and $\sigma_x\otimes\sigma_x + \sigma_y\otimes\sigma_y$ is used for iSWAP gate~\cite{caldwell2018parametrically,arute2019quantum}.

\subsection{Graphs, Cuts, and Odd-Vertex Pairings}

\begin{figure}[!h]
    \centering
    \includegraphics[width=0.9\columnwidth]{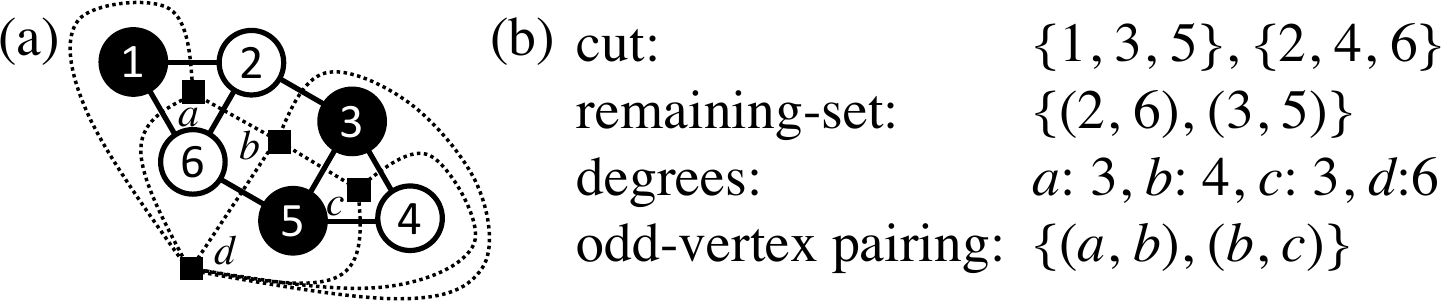}
    \caption{An example of a cut in a planar graph. (The dual graph is drawn by square vertices and dashed edges.)}
    \label{fig:cut}
\end{figure}

The topology of a QC device can be represented by a graph $G=(V,E)$, where $V$ is a vertex set corresponding to qubits, and $E$ is an edge set corresponding to couplings. A planar topology corresponds to a planar graph. Every planar graph $G$ has a \textit{dual graph} $G^*$, where each vertex of $G^*$ corresponds to a face of $G$ (including the outer face), and each edge $e^*$ of $G^*$ corresponds to an edge $e$ of $G$ (see Figure~\ref{fig:cut}~(a)).

A \textit{cut} $C=(S,T)$ is a partition of $V$ into two disjoint sets $S, T$. After removing the edges across the cut, one is left with the \textit{remaining-set} $R_C$ that includes all edges whose endpoints belong to the same partition $R_C = \{(u, v) | u, v \in S \text{ or } u, v \in T\}$ (see Figure~\ref{fig:cut}~(b)).

An edge set whose contraction leaves the remaining graph free of odd-degree vertices is called an \textit{odd-vertex pairing}. (Contracting an edge refers to removing it from a graph and merging the two vertices that it previously joined.) All graphs have an even number of odd-degree vertices, and an odd-vertex pairing covers all the odd-degree vertices. For example, the odd-degree vertices of the dual graph in Figure~\ref{fig:cut} are $a, c$. After contracting the edges $D^* = \{(a, b), (b, c)\}$, $G^*$ will have only two even-degree vertices ($a,b,c$ are merged into one), and thus $D^*$ is an odd-vertex pairing. The following theorem establishes a connection between cuts and odd-vertex pairings.

\begin{theorem}\label{thm:odd-vertex-pairing}
    An edge set $D$ contains a remaining-set $R_C$ for a cut $C$ of a graph $G$ if and only if the dual edge set $D^*$ of $D$ in the dual graph $G^*$ is an odd-vertex pairing of $G^*$.
\end{theorem}

Refer to~\cite{hadlock1975finding,maxcutlecture} for a proof. One can thus find a cut of a graph $G=(V, E)$ as follows. First, find an odd-vertex pairing $D^*$ in the dual graph $G^*$ (e.g., $D^* = \{(a, b), (b, c)\}$). After contracting from $G$ the dual edges $D$ of $D^*$ (e.g., $D = \{(2, 6), (3, 5)\}$), all the remaining edges $E - D$ are across a cut. We can then find the cut by coloring the vertices in the remaining graph using two colors. We will use this method in our scheduling algorithm.


\section{ZZ-Suppressing Objectives for Pulse Optimization}
\label{sec:gate_lvl}

We elaborate on the objective of pulse optimization in this section: to implement quantum gates and also suppress cross-region $ZZ$ crosstalk surrounding the gates. The key result is that we only need to optimize pulses on small systems when taking cross-region crosstalk as the suppression target.

We first discuss basic regions with only one single-qubit or two-qubit gate, and then consider general regions consisting of multiple gates.

\subsection{Basic Region: One Single-Qubit Gate} 

\begin{figure}[!h]
    \centering
    \hspace{-0.1\columnwidth}
    \begin{minipage}[c]{0.3\columnwidth}
        \includegraphics[width=\columnwidth]{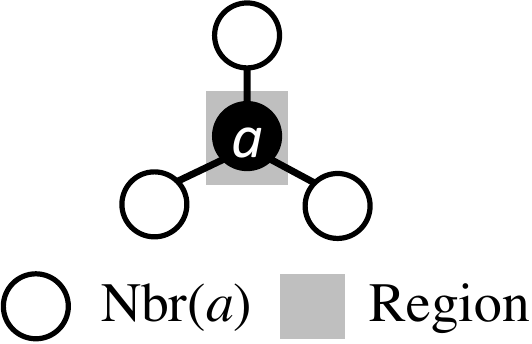}
    \end{minipage}
    \hspace{-2em}
    \begin{minipage}[c]{0.65\columnwidth}
        \centering
        $\begin{aligned}
            H(t) &= \Omega_x^\lb{a}(t) \sigma_x^\lb{a} + \Omega_y^\lb{a}(t) \sigma_y^\lb{a} & (H_{Ctrl}(t))
            \\
            &+ \textstyle{\sum}_{q \in \mathrm{Nbr}(a)} \lambda_{aq} \sigma_z^\lb{a} \otimes \sigma_z^\lb{q} & (\lambda H_{Xtalk})
        \end{aligned}$
    \end{minipage}
    \caption{A basic region with one single-qubit gate $U_1$ on qubit $a$ and its Hamiltonian.}
    \label{fig:single_qubit}
\end{figure}

Figure~\ref{fig:single_qubit} shows a basic region with one single-qubit gate $U_1$ on qubit $a$ and the corresponding Hamiltonian $H(t)$. $\mathrm{Nbr}(a)$ is a set of all the qubits that are neighbors of $a$ and outside the region, and the other notations are defined in Section~\ref{sec:qe}. Using $\lambda = \sum_{q \in \mathrm{Nbr}(a)} \lambda_{aq}$ to denote the total effect of all the cross-region crosstalk, the Hamiltonian can be rewritten as
\begin{equation*}
    H(t) = H_{Ctrl}(t) + \lambda H_{Xtalk}
\end{equation*}
where $H_{Ctrl}(t) = \Omega_x^\lb{a}(t) \sigma_x^\lb{a} + \Omega_y^\lb{a}(t) \sigma_y^\lb{a}$, and $H_{Xtalk} = \sigma_z^\lb{a} \otimes \left( \sum_{q \in \mathrm{Nbr}(a)} \lambda_{aq}\sigma_z^\lb{q} / \lambda \right)$.

We aim to optimize pulses $\Omega_{x,y}^\lb{a}(t)$ to (i) implement $U_1$ on $a$ and (ii) suppress the effect of the cross-region crosstalk $H_{Xtalk}$ surrounding $a$.

The implementation of $U_1$ is achieved by letting $U_{Ctrl}(T) = U_1$ where $U_{Ctrl}(t)$ is defined by $i \hbar \dot{U}_{Ctrl}(t) = H_{Ctrl}(t) U_{Ctrl}(t)$ and $T$ is the duration of the pulses. In other words, the target gate should be implemented when all the crosstalk is suppressed.

To suppress $H_{Xtalk}$, we maximize the similarity between the actual evolution $U(t)$ of the system, defined by $i \hbar \dot{U}(t) = H(t) U(t)$, and the desired evolution $U_1 \otimes \mathbf{I}^{\mathrm{Nbr}(a)}$, where $\mathbf{I}^{\mathrm{Nbr}(a)}$ represents identity operations on $\mathrm{Nbr}(a)$. 

To summarize, the optimization objective is
\begin{equation}
\label{equ:obj-single}
\begin{aligned}
    \underset{\{\Omega(t)\}}{\mathrm{maximize}} & \ \ F\left(U(T), U_1 \otimes \mathbf{I}^{\mathrm{Nbr}(a)}\right) \\
    \text{subject to} & \ \ U_{Ctrl}(T) = U_1
\end{aligned}
\end{equation}
where $\{\Omega(t)\} = \{\Omega_{x,y}^\lb{a}(t)\}$ and $F$ is a measure of similarity.

\subsection{Basic Region: One Two-Qubit Gate} 

\begin{figure}[!h]
    \centering
    \hspace{-0.1\columnwidth}
    \begin{minipage}[c]{0.2\columnwidth}
        \centering
        \includegraphics[width=\columnwidth]{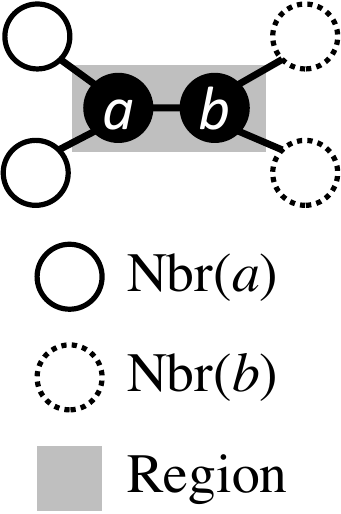}
    \end{minipage}
    \begin{minipage}[c]{0.65\columnwidth}
        \centering
        \scalebox{0.95}{%
        $\begin{aligned}
            H&(t) = \\
            &\left. 
                \arraycolsep=-1pt\begin{array}{l}
                  \sum_{q \in \{a, b\}} \Omega_x^\lb{q}(t) \sigma_x^\lb{q} + \Omega_y^\lb{q}(t) \sigma_y^\lb{q} \\[0.4em]
                  +\Omega^\lb{a\text{-}b}(t)H_{Coupling}
                \end{array}
                \ \right \} (H_{Ctrl}(t)) \\
            &\left. 
              \arraycolsep=-1pt\begin{array}{l}
                +\sum_{q \in \mathrm{Nbr}(a)} \lambda_{aq} \sigma_z^\lb{a} \otimes \sigma_z^\lb{q} \\
                +\sum_{q \in \mathrm{Nbr}(b)} \lambda_{bq} \sigma_z^\lb{b} \otimes \sigma_z^\lb{q}
              \end{array}
              \ \right \} \hspace{2.8em} (\lambda H_{Xtalk}) \\
            &+ \lambda_{ab} \sigma_z^\lb{a} \otimes \sigma_z^\lb{b} \hspace{6em} (\lambda' H_{IntraXtalk})
        \end{aligned}$%
        }
    \end{minipage}
    \caption{A basic region with one two-qubit gate $U_2$ on qubits $a, b$ and its Hamiltonian.}
    \label{fig:two_qubit}
\end{figure}

Figure~\ref{fig:two_qubit} shows a basic region with one two-qubit gate $U_2$ on qubits $a, b$ and its Hamiltonian. The Hamiltonian can be rewritten as
\begin{equation*}
    H(t) = H_{Ctrl}(t) + \lambda H_{Xtalk} + \lambda' H_{IntraXtalk}
\end{equation*}
where $H_{Xtalk}$ denotes crosstalk across regions, and $H_{IntraXtalk}$ denotes crosstalk within the region. 

We aim to optimize pulses to (i) implement $U_2$ on $a, b$ and (ii) suppress the effect of the cross-region crosstalk $H_{Xtalk}$ surrounding $a, b$. Similarly, the optimization objective is
\begin{equation}
\label{equ:obj-two}
\begin{aligned}
    \underset{\{\Omega(t)\}}{\mathrm{maximize}} & \ \ F\left(U(T), \tilde{U}_2(T) \otimes \mathbf{I}^{\mathrm{Nbr}(a)} \otimes \mathbf{I}^{\mathrm{Nbr}(b)} \right) \\
    \text{subject to} & \ \ U_{Ctrl}(T) = U_2
\end{aligned}
\end{equation}
where $\{\Omega(t)\} = \{\Omega_{x,y}^\lb{a}(t),\Omega_{x,y}^\lb{b}(t),\Omega^\lb{a\text{-}b}(t)\}$. The major difference from the basic region with one single-qubit gate is that the desired evolution on $a,b$ is no longer the ideal gate $U_2$ but $\tilde{U}_2(T)$ defined by $i \hbar \dot{\tilde{U}}_2(t) = \left[ H_{Ctrl}(t) + \lambda' H_{IntraXtalk} \right] \tilde{U}_2(t)$, since we allow intra-region crosstalk when optimizing pulses.

\subsection{General Regions}

For a general region composed of multiple single-qubit and two-qubit gates, we do \textit{not} need to optimize again, but can \textit{directly} apply the pulses optimized in basic regions with the same target gates. In this way, all cross-region $ZZ$ crosstalk for that general region can be suppressed simultaneously.

\begin{figure}[ht]
    \centering
    \begin{minipage}[c]{0.2\columnwidth}
        \centering
        \includegraphics[width=0.9\columnwidth]{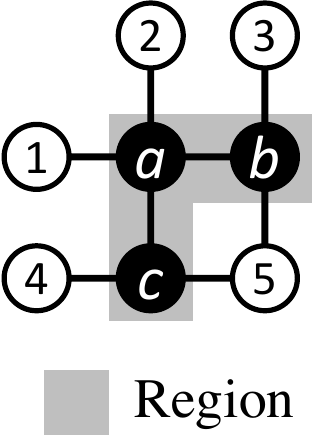}
    \end{minipage}
    \begin{minipage}[c]{0.7\columnwidth}
        \centering
        \scalebox{0.85}{%
        $\begin{aligned}
            H(t) &= H_{Ctrl}^\lb{c}(t) + \sum_{q \in \{4, 5\}} \lambda_{cq} \sigma_z^\lb{c} \otimes \sigma_z^\lb{q} \\
            &+ H_{Ctrl}^\lb{a, b}(t) + \sum_{q \in \{1, 2\}} \lambda_{aq} \sigma_z^\lb{a} \otimes \sigma_z^\lb{q} \\
            &\hspace{4.2em} + \sum_{q \in \{3, 5\}} \lambda_{bq} \sigma_z^\lb{b} \otimes \sigma_z^\lb{q}\\
            &+ \lambda' H_{IntraXtalk}
        \end{aligned}$%
        }
    \end{minipage}
    \caption{A general region with a two-qubit gate $U_2$ on qubits $a, b$ and a single-qubit gate $U_1$ on qubit $c$.}
    \label{fig:many_qubit}
\end{figure}

For instance, consider pulses optimized in two basic regions, one with only a single-qubit gate $U_1$ and the other with a two-qubit gate $U_2$. Then, for the general region in Figure~\ref{fig:many_qubit}, by directly applying these optimized pulses, the pulse for $U_1$ on qubit $c$ can suppress $c$-4 and $c$-5, and the pulse for $U_2$ on qubits $a, b$ can suppress $a$-1, $a$-2, $b$-3, $b$-5. Thus, the application of these two pulses suppresses all of the crosstalk across regions. One can find out, in a similar manner, that this result holds for arbitrary general regions. The behind reason is that each cross-region coupling is connected to one qubit with a gate performed on, and we can suppress crosstalk on that coupling by the pulses for that gate.

The above result implies \textit{high scalability} in that we only need to optimize pulses \textit{in basic regions} (i.e., small systems), which avoids the high time and memory consumption of optimization in general regions.

In Section~\ref{sec:evaluation}, we will give three approaches to implement the optimization objectives for basic regions: quantum optimal control, dynamic corrected gates and a new proposed one based on quantum perturbative theory.


\newcommand{\optsuppression}{
\SetKwRepeat{Do}{repeat}{until}
\begin{algorithm}[t]
        \DontPrintSemicolon
        \caption{$\alpha$-Optimal Suppression Algorithm}
        \label{algo:optimal}
        \KwIn{Topology graph $G=(V,E)$; Qubits involved in gates $Q$; Relative importance coefficient $\alpha$}
        \KwOut{A cut for $\alpha$-optimal suppression}

        $G^* \gets$ the dual graph of $G$ \;
        $E_{Q}^* \gets \{(u,v)^*|(u,v) \in E \ \mathrm{and}\ u,v \in Q\}$ \;
        \textbf{\underline{Delete Edges:}} $G^* \gets G^* - E_Q^*$ \;
        \SetKwBlock{VP}{\textnormal{\textbf{\underline{Vertex Pairing:}}}}{}
        \VP{
            $G_{odd} \gets$ a complete graph with the odd-degree vertices of $G^*$ and edge weights $w(u,v) = L - d(u,v)$ \;
            $M \gets$ a maximum matching of $G_{odd}$ \;
        }
        \SetKwBlock{PR}{\textnormal{\textbf{\underline{Path Relaxing:}}}}{}
        \SetKwBlock{CK}{\textnormal{\textbf{\underline{Check:}}}}{}
        \SetKwBlock{CI}{\textnormal{\textbf{\underline{Cut Inducing:}}}}{}
        \PR{ \label{line:pr_begin}
            \ForEach{pair of matched vertices $(u, v) \in M$}{
                $L^{\lb{u,v}} \gets [L_1^{\lb{u,v}}, L_2^{\lb{u,v}}, \cdots]$ a list of top-$k$ shortest paths between $(u,v)$ sorted by their length in ascending order \;
            }
            Odd-vertex pairing $P \gets \{L_1^{\lb{u,v}}|(u,v) \in M\}$ \;
            \Do{$\alpha N_Q + N_C$ of $P$ is unchanged}{
                $candidates \gets \text{\O} $ \;
                \ForEach{$(u, v) \in M$}{ \label{line:check}
                    $P' \gets P - L_i^{\lb{u,v}} + L_{i+1}^{\lb{u,v}}$ \quad \textit{// relax} $L_i^{\lb{u,v}} = P \cap L^{\lb{u,v}}$ \;
                    \CI{
                        \textbf{\underline{Add Edges:}} $P' \gets P' + E_Q^*$ \;
                        $G' \gets G\text{ with the dual edges of }P'\text{ contracted}$ \;
                        Obtain a cut $C$ by coloring the vertices of $G'$ with 2 colors\;
                    }
                    \textbf{\underline{Check:}} \lIf{$Q \subset$ a partition of $C$}{add $P'$ to $candidates$}
                }
                $P \gets$ the candidate of $candidates$ with minimum $\alpha N_Q + N_C$ \; \label{line:update}
            }
        }\label{line:pr_end}
        
        \Return{a cut $C$ induced from $P$}
\end{algorithm}
}

\section{Optimal Suppression}
\label{sec:layer_lvl}

In this section we show how a layer of gates can be executed, assisted by extra identity gates, such that $ZZ$ crosstalk on the whole device is maximally suppressed.

The key idea is to establish a connection between the status of qubits and a cut of the device topology. A qubit's status is either (i) with a gate (pulse) applied to it or (ii) not. The qubits with gates acting on them form a set $S$, the others constitute a set $T$, and these sets form a cut $C = (S, T)$ of the topology. Given a layer $L$ of gates, the aim is to find such a cut $(S,T)$ that (i) the qubits acted on by gates in $L$ all lie in $S$, and (ii) any remaining qubits in $S$ are acted on by identity gates to further improve crosstalk suppression.

Before discussing the general case where the layer contains multiple gates, we first consider a special case where it contains no gates, and one simply wishes to suppress crosstalk by applying identity gates to all the qubits in $S$.

\subsection{Layers Containing No Gates}

\begin{figure}[!h]
    \centering
    \includegraphics[width=0.6\columnwidth]{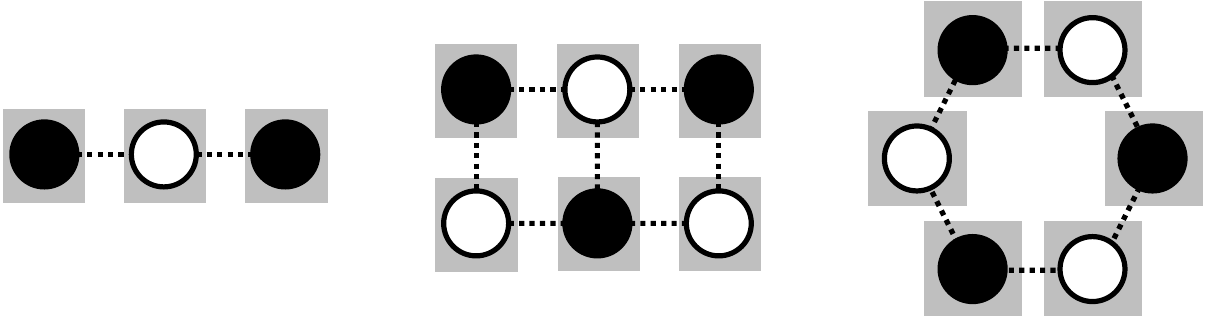}
    \caption{Examples of complete suppression. All crosstalk is suppressed by applying identity pulses to black qubits.}
    \label{fig:complete}
\end{figure}

Ideally, all the crosstalk on a device can be suppressed by applying identity pulses to certain qubits. We refer to this as \textit{complete suppression}. It can be easily proved that complete suppression can be achieved on any device of a bipartite topology (e.g., Figure~\ref{fig:complete}). Fortunately, most near-term QC devices~\cite{murali2019full,ibmq} and topological QEC codes~\cite{campbell2017roads,fowler2012surface} indeed have a bipartite topology.

For devices that do not admit complete suppression, there is at least one region with unsuppressed internal crosstalk. In this case, two metrics are important: \#qubits in the largest region ($N_Q$) and \#couplings with unsuppressed crosstalk ($N_C$) corresponding to all the edges within regions. There is a trade-off between the two metrics, as can be seen in Figure~\ref{fig:optimal_plan}. To characterize this trade-off, we define optimal suppression or, more precisely, \textit{$\alpha$-optimal suppression}.

\def\bigex{3}

\begin{figure}[!t]
    \centering
    \if\bigex1%
        \includegraphics[width=\columnwidth]{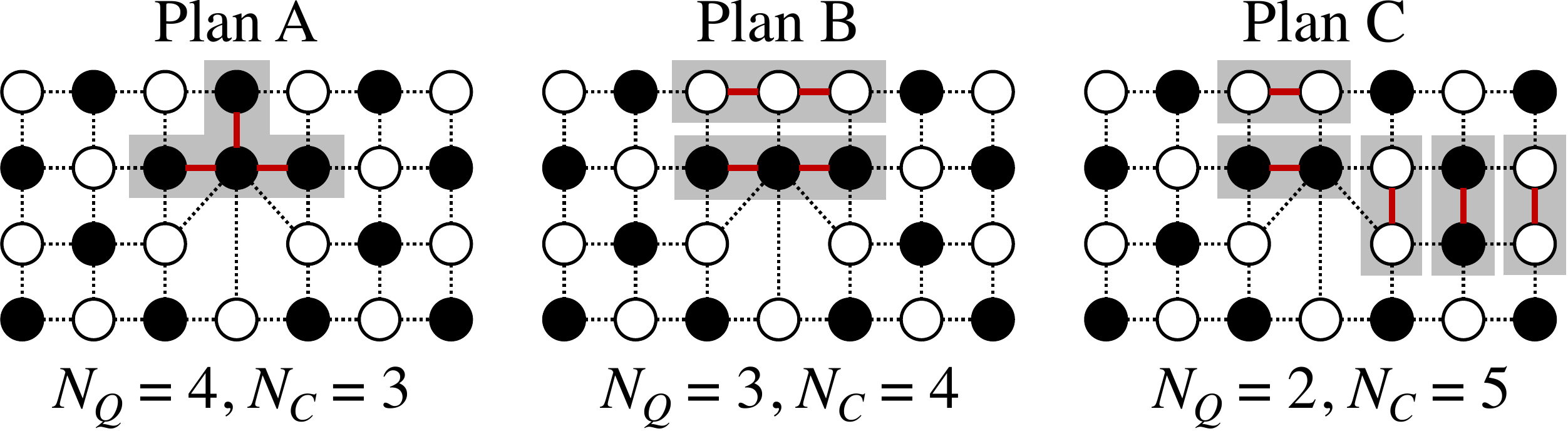}
    \else%
        \if\bigex2%
            \includegraphics[width=0.7\columnwidth]{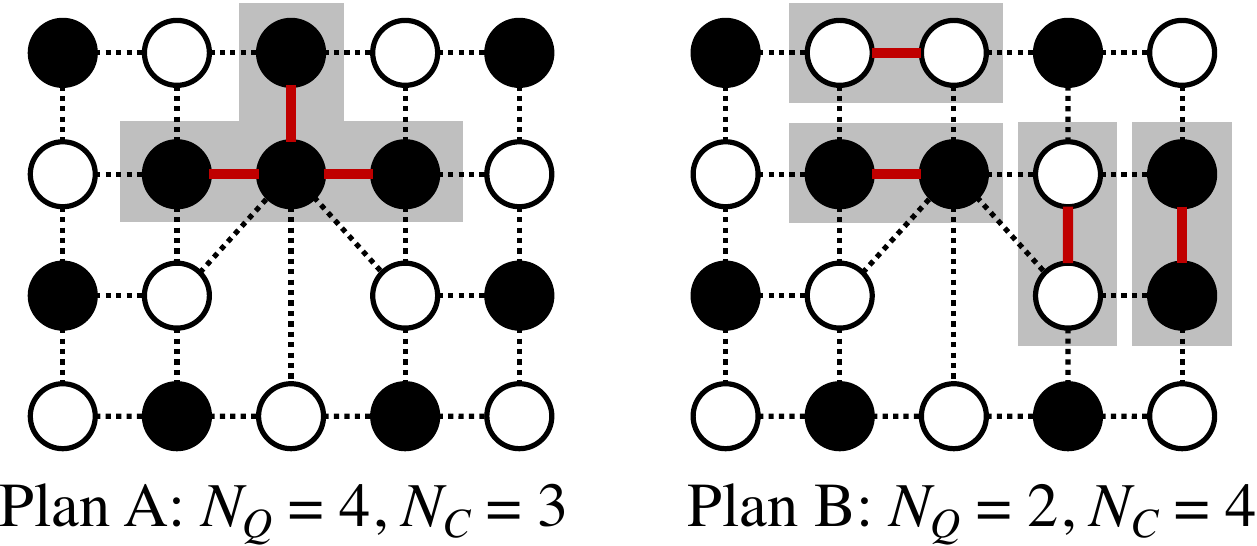}
        \else%
            \includegraphics[width=\columnwidth]{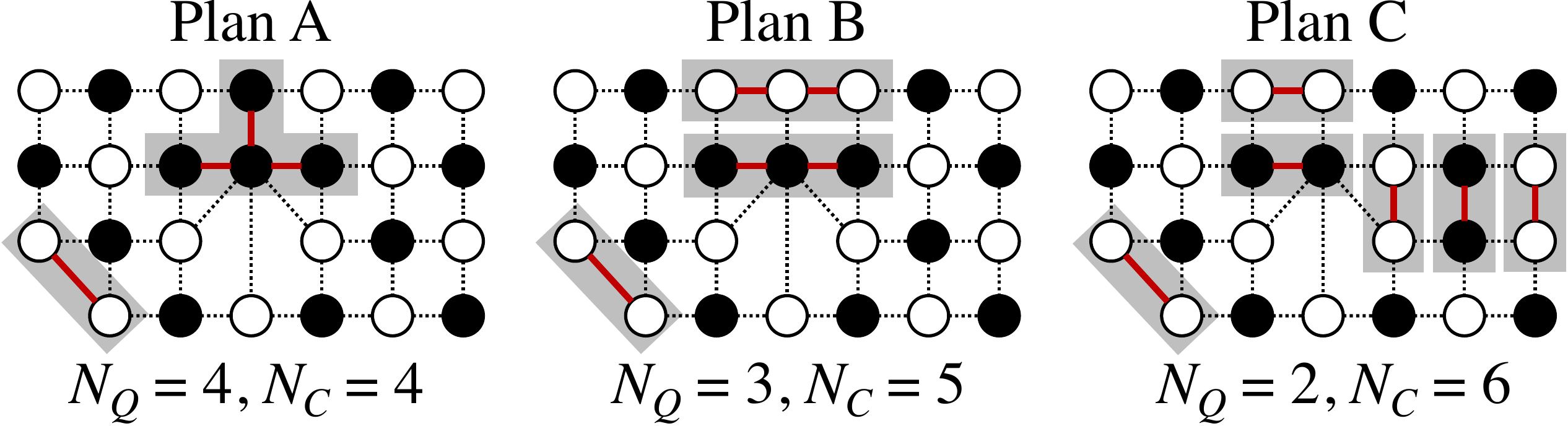}
        \fi
    \fi
    \caption{The trade-off between $N_Q$ and $N_C$.}
    \label{fig:optimal_plan}
\end{figure}

\begin{definition}[$\alpha$-Optimal Suppression]
    Given $\alpha > 0$, a device has an $\alpha$-optimal suppression plan if applying pulses to some of its qubits divides its qubits into multiple regions such that $\alpha N_Q + N_C$ is minimized.
\end{definition}

To find an $\alpha$-optimal suppression plan, we convert the problem to a cut problem of topology graphs.

\begin{note}[A Cut Problem for $\alpha$-Optimal Suppression.]
    Given a graph $G = (V, E)$ and $\alpha > 0$, find a cut $C=(S,T)$ of $G$ that minimizes $\alpha N_Q + N_C$, where $N_Q$ is \#vertices in the largest connected component of the graph $G' = (V, R_C)$, $N_C = |R_C|$, and $R_C$ is the remaining-set of the cut $C$.
\end{note}

The two partitions $S,T$ represent qubits with and without pulses applied to respectively. The remaining-set $R_C$ represents couplings with unsuppressed crosstalk since the endpoints of any edge in $R_C$ belong to the same partition. After giving a solution to the cut problem, $\alpha$-optimal suppression can be achieved by applying identity pulses to the partition $S$. In Figure~\ref{fig:optimal_plan}, the black and white vertices correspond to the partitions $S, T$ respectively, solid edges correspond to the remaining-set $R_C$, and crosstalk on the solid edges is unsuppressed.

The cut problem is difficult to solve. For example, for $\alpha = 0$, we need to find a cut that has a minimum remaining-set $R_C$, i.e., a maximum number of cross-cut edges. This is the maximum cut problem, and is NP-complete for general graphs. Noting that the topology of most near-term QC devices~\cite{murali2019full,ibmq} and topological QEC codes~\cite{campbell2017roads,fowler2012surface} is planar, we therefore aim to design an efficient algorithm for planar topologies.

The main idea of our algorithm is as follows. Based on Theorem~\ref{thm:odd-vertex-pairing}, we can find cuts by locating odd-vertex pairings in dual graphs. Since an odd-vertex pairing covers all odd-degree vertices, the smallest one contains only simple paths, with each path connecting two odd-degree vertices~\cite{hadlock1975finding}. By selecting shortest paths, one can minimize the size of an odd-vertex pairing. As a consequence of Theorem~\ref{thm:odd-vertex-pairing}, $N_C = |R_C|$ can also be minimized. To account for $N_Q$, we consider the top-$k$ shortest paths, and select the path which strikes the desired balance between $N_C$ and $N_Q$.

Specially, our algorithm consists of three steps (see Figure~\ref{fig:optimal_algo} for an example):

\begin{figure}[!t]
    \centering
    \if\bigex1%
        \includegraphics[width=\columnwidth]{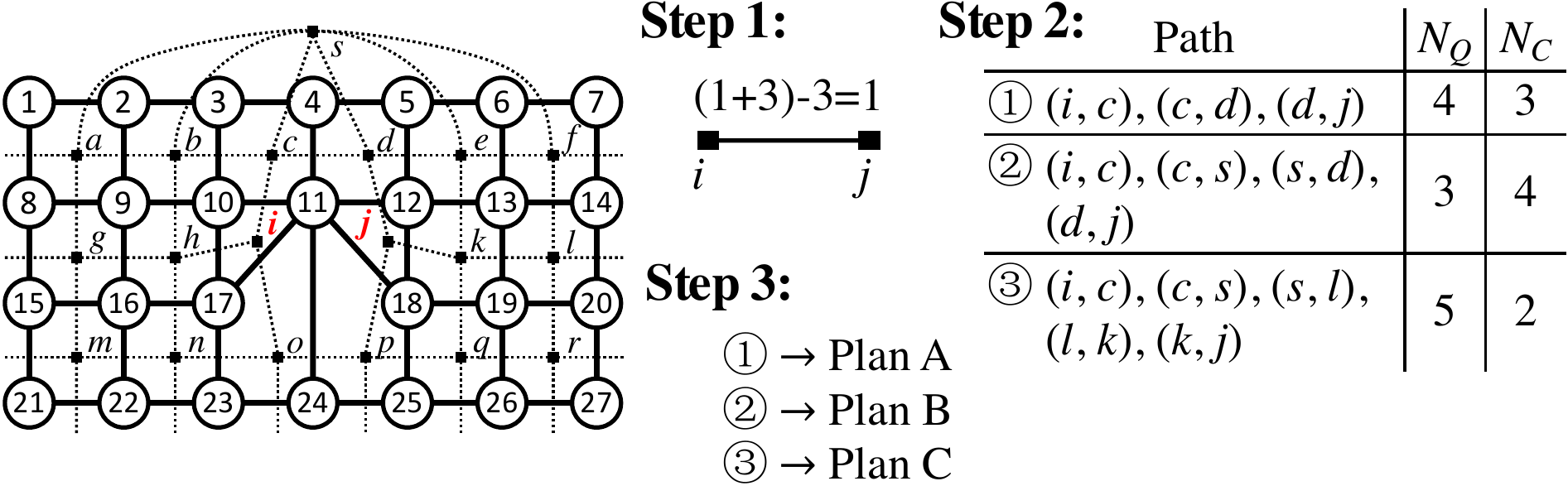}
    \else%
        \if\bigex2
            \includegraphics[width=0.95\columnwidth]{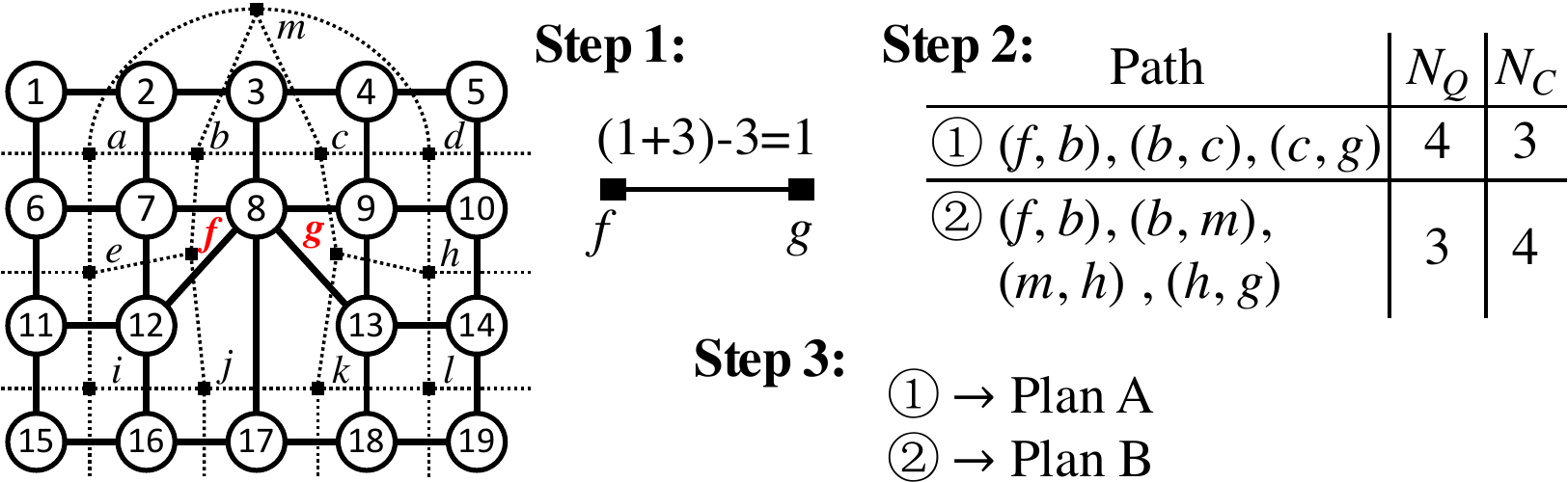}
        \else
            \includegraphics[width=\columnwidth]{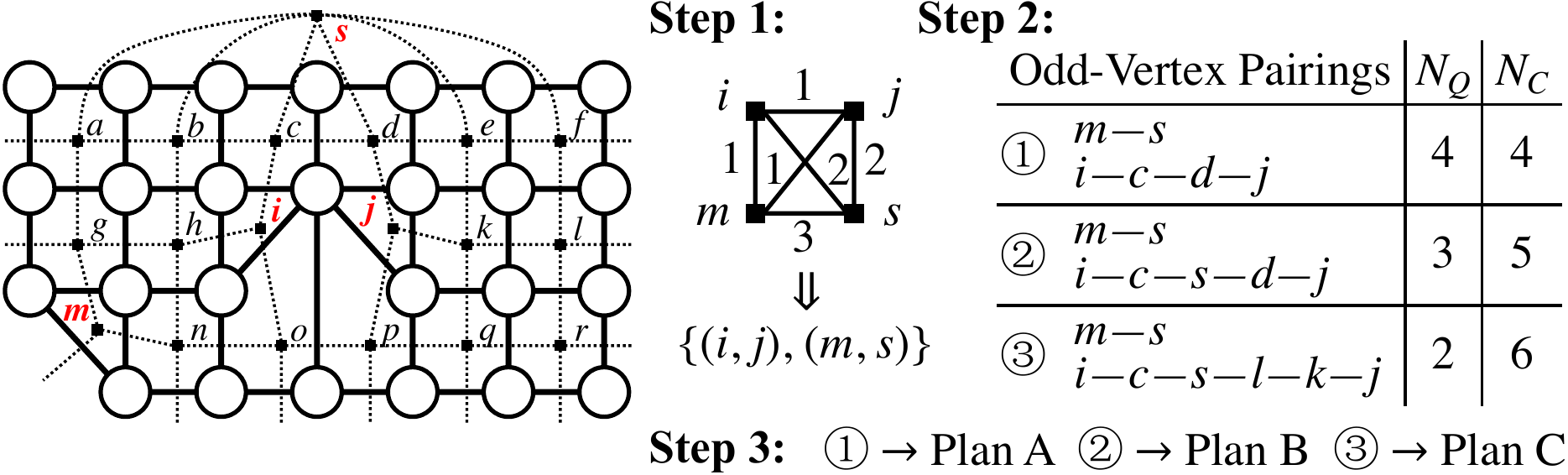}
        \fi
    \fi
    \caption{Left: The topology graph (circular vertices, solid edges) of Figure~\ref{fig:optimal_plan} and its dual (square vertices, dashed edges, all edges on the border are connected to vertex \if\bigex1$s$\else\if\bigex2$m$\else$s$\fi\fi). Right: Steps for running our algorithm on the left topology.}
    \label{fig:optimal_algo}
\end{figure}

\textbf{Step 1: Vertex Matching.} This step decides which two vertices each path should connect. First, we construct a complete graph containing all the odd-degree vertices of the dual graph. Then, a weight $L - d(u,v)$ is assigned to each edge $(u,v)$, where $d(u, v)$ is the shortest path length between $u,v$, and $L = 1 + \max_{u,v} d(u,v)$. Finally, the vertices on the complete graph are matched by a maximum weight matching algorithm~\cite{galil1986efficient}. In this way, connecting each pair of matched vertices in the dual graph by shortest paths can give a smallest odd-vertex pairing~\cite{hadlock1975finding}.%
\if\bigex1%
\ In Figure~\ref{fig:optimal_algo}, we construct a complete graph of odd-degree vertices $i, j$. Since $d(i,j) = 3$, the weight for $(i,j)$ is $(1 + 3) - 3 = 1$. And the maximum matching is $\{(i,j)\}$.
\else\if\bigex2%
\ In Figure~\ref{fig:optimal_algo}, we construct a complete graph of odd-degree vertices $f, g$. Since $d(f,g) = 3$, the weight for $(f,g)$ is $(1 + 3) - 3 = 1$. And the maximum matching is $\{(f,g)\}$.
\else
\ In Figure~\ref{fig:optimal_algo}, we construct a complete graph of odd-degree vertices $\{i, j, m, s\}$, whose maximum weight matching is $\{(i,j), (m,s)\}$.
\fi
\fi

\textbf{Step 2: Path Relaxing.} This step relaxes the shortest path restriction, to enable optimizing over both $N_Q$ and $N_C$. We generate many odd-vertex pairings by connecting each pair of matched vertices with their top-$k$ shortest paths, and then select the one with minimum $\alpha N_Q + N_C$. In more detail, we use a greedy strategy as shown in Algorithm~\ref{algo:optimal}, Line~\ref{line:pr_begin}-\ref{line:pr_end}.%
\if\bigex1%
\ In Figure~\ref{fig:optimal_algo}, we list the top-3 shortest paths between $i,j$ with their $N_Q$ and $N_C$.
\else\if\bigex2%
\ In Figure~\ref{fig:optimal_algo}, we list the top-2 shortest paths between $f,g$ with their $N_Q$ and $N_C$.
\else%
\ In Figure~\ref{fig:optimal_algo}, we generate three samples corresponding to the shortest path between $m,s$ and top-3 shortest paths between $i,j$.
\fi
\fi
We select one according to $\alpha$.

\textbf{Step 3: Cut Inducing.} This step obtains a cut for the selected odd-vertex pairing. Based on Theorem~\ref{thm:odd-vertex-pairing}, after contracting the dual edges of the selected odd-vertex pairing in the topology graph, we obtain a cut by performing breadth-first search to color the vertices of the remaining topology graph using 2 colors.%
\if\bigex1%
\ In Figure~\ref{fig:optimal_algo}, the odd-vertex pairings \C1, \C2, \C3 obtained from step 2 correspond to Plan A, B, C in Figure~\ref{fig:optimal_plan} respectively.
\else\if\bigex2%
\ In Figure~\ref{fig:optimal_algo}, the odd-vertex pairings \C1 and \C2 obtained from step 2 correspond to Plan A and B in Figure~\ref{fig:optimal_plan} respectively.
\else%
\ In Figure~\ref{fig:optimal_algo}, the odd-vertex pairings \C1, \C2, \C3 obtained from step 2 correspond to Plans A, B, C in Figure~\ref{fig:optimal_plan} respectively.
\fi
\fi

\subsection{Layers Containing Gates}

\optsuppression

This case can be regarded as a constrained problem that requires all the qubits involved in the gates of a layer belong to the same partition, because we need to apply pulses to all of them. We solve this case by converting it to the unconstrained problem (layers containing no gates). The conversion is based on a property of remaining-sets.

\begin{theorem}
    The remaining-set of a cut comprises multiple connected components, with vertices in the same connected component belonging to the same partition of the cut.
\end{theorem}

For example, gray areas in Figure~\ref{fig:optimal_plan} identify connected components, and the vertices in the same gray area have the same color, i.e., they belong to the same partition.

Based on this theorem, denoting the set of qubits involved in the gates of a layer by $Q$ and the edges involved $\{(u,v)|(u,v) \in E\allowbreak \ \mathrm{and}\ \allowbreak u,v \in Q\}$ by $E_Q$, if we explicitly add $E_Q$ to a remaining-set, then the vertices of $Q$ in the same connected component of graph $(Q, E_Q)$ necessarily belong to the same partition of the induced cut. Then we only need to check whether the vertices in different connected components belong to the same partition. Specifically, we perform the conversion in the dual graph in three steps. We denote the dual of $E_Q$ by $E_Q^*$.

\textbf{Step 1: Delete Edges.} Delete $E_Q^*$ from the dual graph. Then preform Vertex Matching and Path Relaxing on the remaining dual graph.

\textbf{Step 2: Add Edges.} Add $E_Q^*$ back to the odd-vertex pairing obtained from Path Relaxing.

\textbf{Step 3: Check.} Perform Cut Inducing according to the modified odd-vertex pairing, and check whether all the vertices of $Q$ belong to the same partition of the induced cut.

\begin{figure}[!t]
    \centering
    \includegraphics[width=\columnwidth]{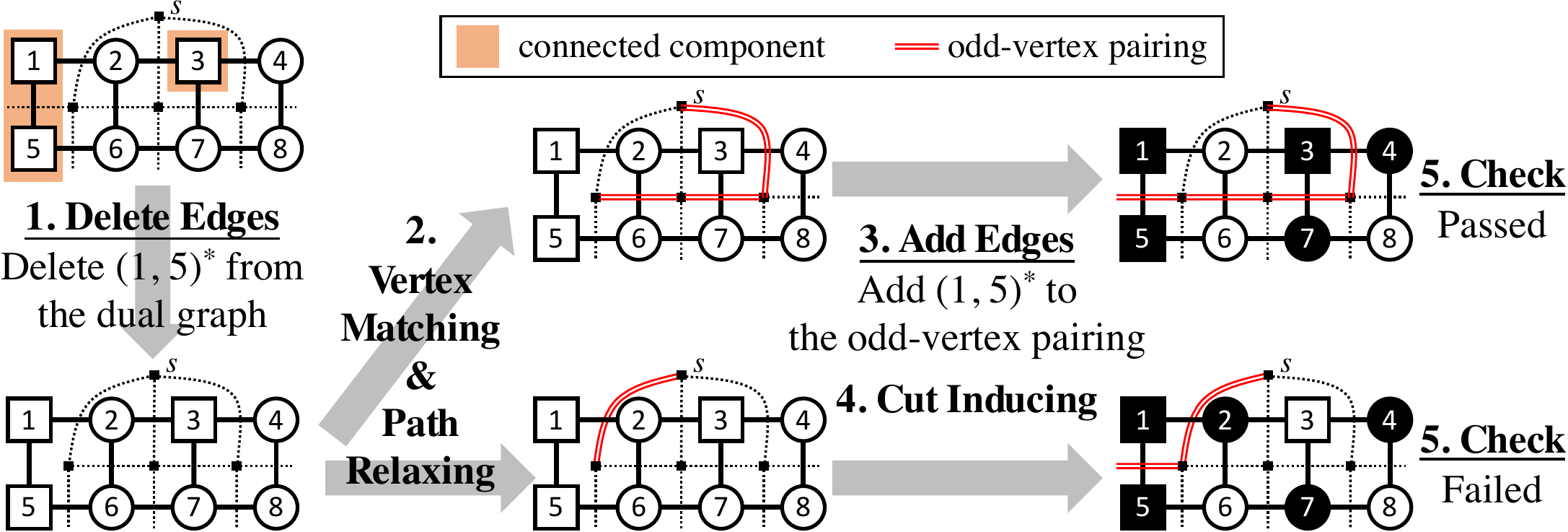}
    \caption{Steps for running our algorithm to perform gates on qubits 1, 3 and 5. (All edges on the border are connected to vertex $s$.)}
    \label{fig:optimal_gates}
\end{figure}

For the example in Figure~\ref{fig:optimal_gates}, $Q=\{1, 3, 5\}$ and $E_Q=\{(1,5)^*\}$, graph $(Q, E_Q)$ has two connected components. We first delete the edge $(1, 5)^*$ from the dual graph. After obtaining odd-vertex pairings, we add $(1,5)^*$, induce a cut and check the constraint. The example shows two odd-vertex pairings, one passes the check while the other does not.

\myparagraph{Time Complexity Analysis} We summarize the complete process in Algorithm~\ref{algo:optimal}. We use the blossom algorithm~\cite{galil1986efficient} for maximum weight matching in $O(n_o^3)$ time, where $n_o$ is \#odd-degree vertices. We use Yen's algorithm~\cite{yen1970algorithm} to generate top-$k$ shortest paths in $O(kn_d(m_d+n_d\log n_d))$ time, where $n_d$ is \#vertices in the dual graph and $m_d$ is \#edges. For path relaxing, our greedy strategy takes $O(kn_o/2)$ iterations, since each path can be used in Line~\ref{line:update} at most once. In each iteration, our strategy checks $O(n_o/2)$ paths (Line~\ref{line:check}). For cut inducing, breadth-first search takes $O(n_t + m_t)$ time, where $n_t$ is \#vertices in the topology graph and $m_t$ is \#edges. In total, our algorithm takes $O(kn^3 \log n)$ time, where $n = \max\{n_t, m_t, n_d\}$.


\section{ZZ-Aware Scheduling}
\label{sec:circuit_lvl}

\newcommand{\scheduling}{
\begin{algorithm}[t]
        \DontPrintSemicolon
        \SetKwProg{subproc}{Procedure}{}{}
        \caption{Schedule a Quantum Circuit}
        \label{algo:circuit_schedule}
        \KwIn{Topology graph $G$; Quantum circuit $QC$; Suppression requirement $R$; $\alpha$ for the  $\alpha$-optimal suppression algorithm}
        \KwOut{A scheduling plan: layers of simultaneous gates}

        \While{there exists unscheduled gates of $QC$}{
            $SG \gets$ a set of all schedulable gates \; \label{line:exe}
            \eIf(\tcp*[f]{Case 1}){$SG$ has only single-qubit gates}{ \label{line:type_start}
                Cut $(S, T) \gets$ $\alpha$-optimal suppression with topology $G$ and qubits \O{} ($S$ has more qubits involved in $SG$ than $T$) \; \label{line:single}
            }(\tcp*[f]{Case 2: $SG$ contains two-qubit gates}){
                $SG_2 \gets$ two-qubit gates of $SG$ \; \label{line:two}
                Partition $S \gets$ \textsc{TwoQSchedule}($G$, $SG_2$, $R$) \;
            } \label{line:type_end}
            \textsc{Schedule}($S$, $SG$) \; \label{line:schedule}
        }

        \subproc{\textnormal{\textsc{Schedule}(Partition $S$, Gates $SG$)}}{ \label{line:subproc_schedule}
            $Q \gets$ qubits involved in gates $SG$ \;
            Layer $L = \{\text{gate }g \in SG~|~\text{all qubits for gate }g \in S \}$ \;
            \ForEach(\tcp*[f]{supplement $L$ with identity gates}){qubit $q \in S - Q$}{ \label{line:add_id}
                $L = L \cup\{$an identity gate on $q\}$
            }
            \textbf{yield} $L$ \;
        }

        \subproc{\textnormal{\textsc{TwoQSchedule}(Topology $G$, Two-qubit gates $SG_2$, Suppression requirement $R$)}}{ \label{line:subproc_two}
            Cut $C = (S,T) \gets$ $\alpha$-optimal suppression with topology $G$ and qubits involved in $SG_2$ ($S$ contains qubits involved in $SG_2$) \;
            \lIf{$C$ satisfies $R$}{
                \Return{$S$}\tcp*[f]{schedule simultaneously}
            }
            \Else(\tcp*[f]{heuristic selection according to distance}){
                $a, b \gets $ two gates of $SG_2$ with the minimum distance \; \label{line:min_distance_gate}
                Groups $A \gets \{a\}, B \gets \{b\}$ \; \label{line:group}
                \While{$SG_2$ is not empty}{ \label{line:distance_start}
                    $(M_g, M_G) \gets$ the gate (of $SG_2$) and group ($A$ or $B$) with the maximum distance \; \label{line:max_distance_gate}
                
                    Cut $C \gets$ $\alpha$-optimal suppression with topology $G$ and qubits involved in $M_G$ + $M_g$ \;
                    \lIf{$C$ satisfies $R$}{$M_G$ += $M_g$, $SG_2$ -= $M_g$}
                    \lElse{\textbf{break}} \label{line:break}
                } \label{line:distance_end}
                \leIf{$|A| > |B|$}{$M \gets A$}{$M \gets B$} \label{line:more}
                Cut $(S, T) \gets$ $\alpha$-optimal suppression with topology $G$ and qubits involved in $M$ ($S$ contains qubits involved in $M$) \;
                \Return{$S$}
            }
        }
\end{algorithm}
}

We present a complete scheduling algorithm in this section, with the focus on how to partition a circuit into multiple layers so as to strike a balance between crosstalk suppression and parallelism.

\scheduling

Our scheduling algorithm is given in Algorithm~\ref{algo:circuit_schedule}. The main idea is to iteratively schedule the gates in schedulable gate sets\footnote{A gate is schedulable if all of its preceding gates have been scheduled, and a schedulable gate set consists of all currently schedulable gates.} by making crosstalk suppression the first priority and then maximizing parallelism. Given a schedulable gate set, two cases can arise: (1) it contains only single-qubit gates, or (2) it contains two-qubit gates. In the first case, we propose a strategy that can always achieve complete suppression for bipartite topologies. In the second case, we propose a heuristic based on the distance between two-qubit gates.

\begin{figure}[!h]
    \centering
    \includegraphics[width=0.95\columnwidth]{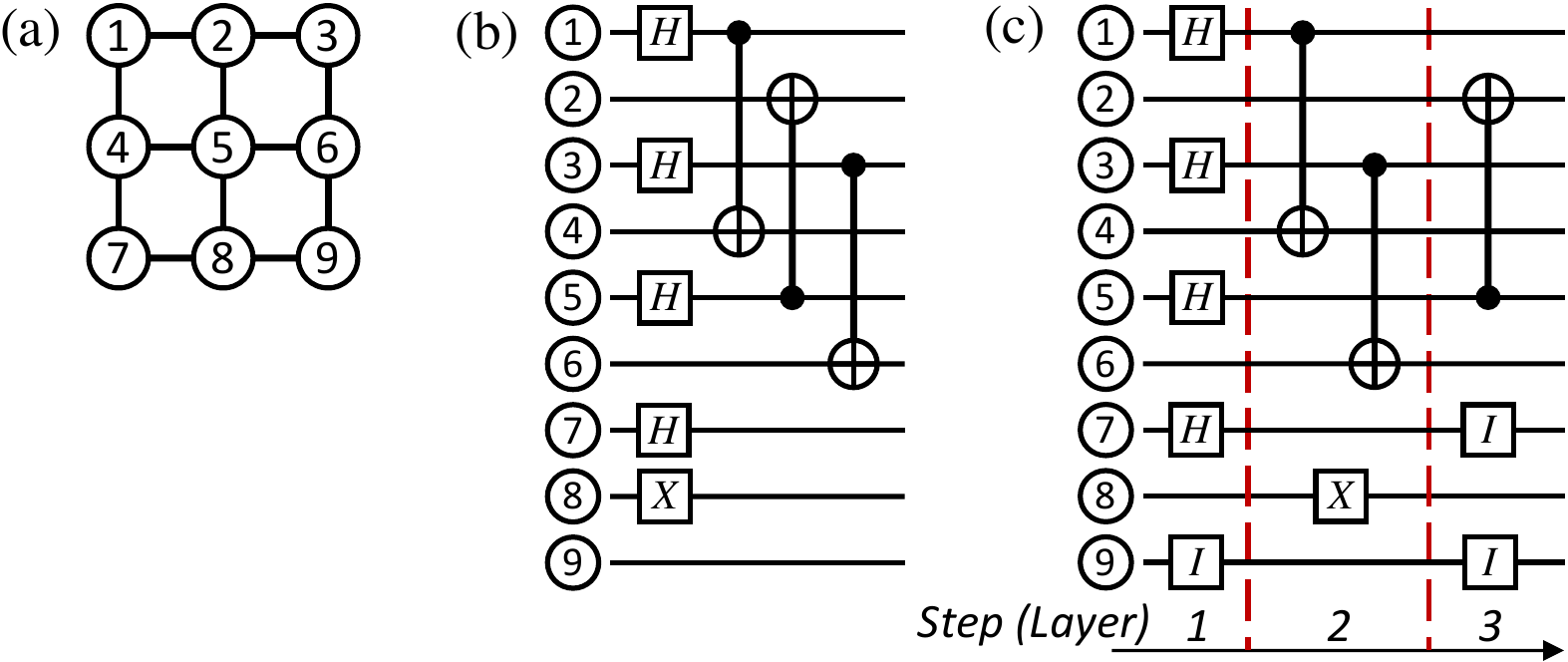}
    \caption{An example for scheduling the quantum circuit (b) on the device (a). Our scheduling plan is shown in (c). (Assume the same duration for all the gates, different durations will be considered in the evaluation.)}
    \label{fig:schedule_example}
\end{figure}

Figure~\ref{fig:schedule_example} shows an example. When no gates have been scheduled, the schedulable gate set is $\{H_{1,3,5,7},X_8\}$ (Case 1). After scheduling layer 1, the set will be $\{\CNOT_{1,4},\allowbreak \CNOT_{5,2},\allowbreak \CNOT_{3,6},\allowbreak X_8\}$ (Case 2).

\myparagraph{Case 1. Only single-qubit gates} The strategy for this case is based on an observation about bipartite topologies.

\myleading{Observation} Complete suppression can be achieved on bipartite topologies with single-qubit gates. (This is illustrated by Figure~\ref{fig:complete} in Section~\ref{sec:layer_lvl}.)

\begin{figure}[!t]
    \centering
    \includegraphics[width=0.5\columnwidth]{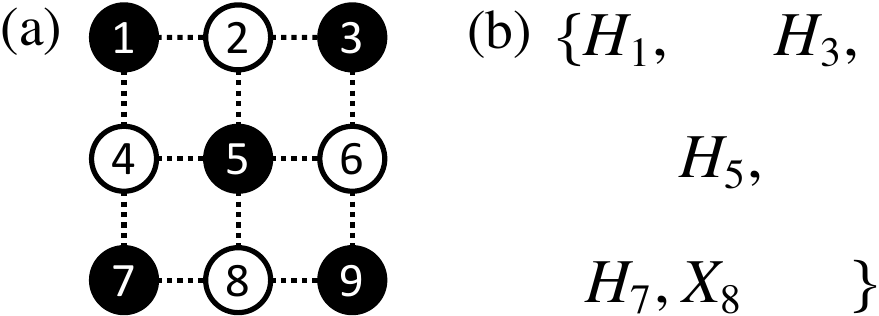}
    \caption{(a) The suppression plan given by the optimal suppression algorithm. (b) The schedulable gates for Layer 1.}
    \label{fig:schedule_no_gates}
\end{figure}

\myleading{Strategy} We thus first perform the $\alpha$-optimal suppression algorithm for layers containing no gates (Line~\ref{line:single}), which, for bipartite topologies, gives the cut (partitions of qubits) that can admit complete suppression (Figure~\ref{fig:schedule_no_gates}~(a)). According to the qubits they act on, the schedulable gates are also divided into two partitions (Figure~\ref{fig:schedule_no_gates}~(b)). Then, we schedule the partition that has more gates to maximize the parallelism, i.e., grouping these gates as a layer and supplementing it with additional identity gates (see Procedure \textsc{Schedule}). In Figure~\ref{fig:schedule_no_gates}, we choose the gates on black qubits to schedule, with an identity gate on qubit $9$. The remaining gates (e.g., $X_8$) will be scheduled in the following layers.

\hyphenation{Two-Q-Schedule}

\myfirstparagraph{Case 2. The schedulable gate set contains two-qubit gates} The strategy for this case is based on an observation about the distance between two-qubit gates.

\begin{definition}
    The distance between two-qubit gates $a$ on qubits $a_1, a_2$ and $b$ on qubits $b_1, b_2$ is defined by $D(a, b) = \sum_{i,j} d(a_i, b_j)$ where $d(a_i, b_j)$ is the length of the shortest path between qubits $a_i$ and $b_j$.
\end{definition}
\begin{definition}
    The distance between a two-qubit gate $a$ and a group $G$ of two-qubit gates is defined by $D(a, G) = \min_{g \in G} D(a, g)$.
\end{definition}

\myleading{Observation} Executing two-qubit gates with a shorter distance usually worsens suppression.

\begin{figure}[!h]
    \centering
    \includegraphics[width=\columnwidth]{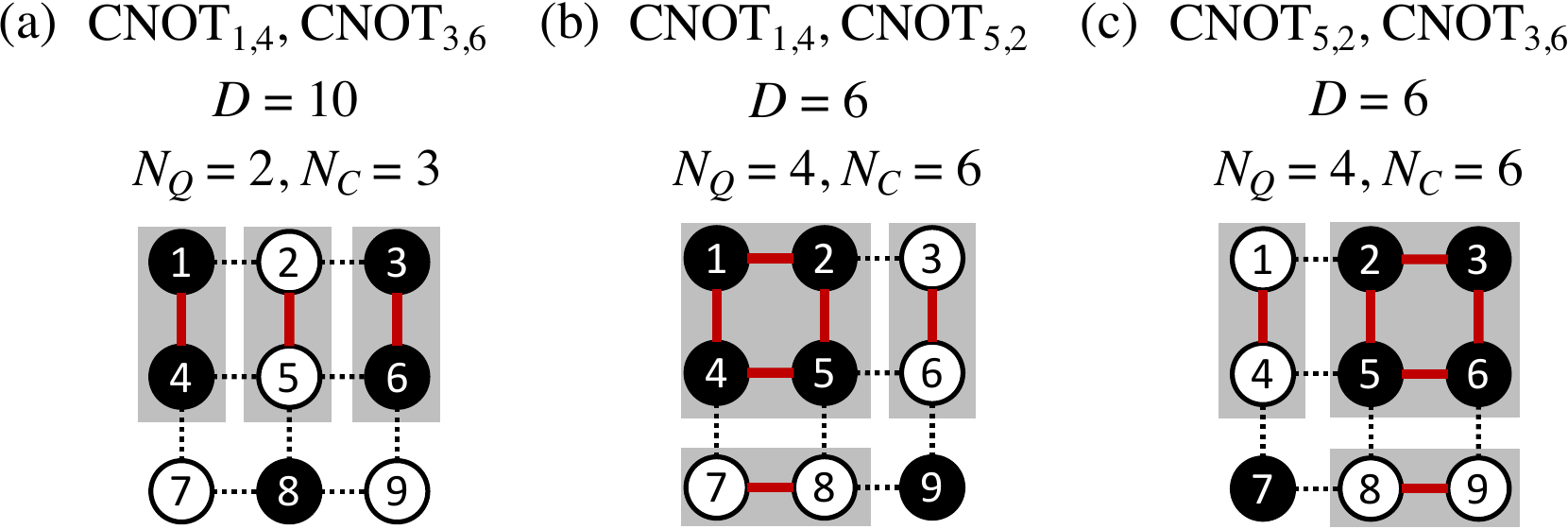}
    \caption{Suppression performance ($N_Q$ and $N_C$, lower is better) and distances ($D$) of different ways to execute two CNOT gates simultaneously.}
    \label{fig:schedule_distance}
\end{figure}

Figure~\ref{fig:schedule_distance} compares the suppression performance and the distances of different ways to execute the CNOT gates in Figure~\ref{fig:schedule_example}. Executing $\CNOT_{1,4}$ and $\CNOT_{3,6}$ in parallel has a better suppression performance than the other ways, and also corresponds to a larger distance between gates.

\myleading{Strategy} The main idea is to iteratively select the next farthest gate so that we can schedule as many gates as possible to maximize parallelism before a suppression requirement is violated (Procedure \textsc{TwoQSchedule}). Here we define the suppression requirement by some constraints on $N_Q$ and $N_C$ (e.g., requiring them to be less than certain thresholds), and violating the requirement means the suppression performance is too bad to be accepted. 

Specifically, we first separate two closest gates into different groups (Line~\ref{line:min_distance_gate},~\ref{line:group}) and then fill the two groups with the next farthest gate iteratively (Line~\ref{line:distance_start}-\ref{line:distance_end}). For the 3 CNOT gates in Figure~\ref{fig:schedule_example}, we first separate the two closest gates $\CNOT_{1,4}$ and $\CNOT_{5,2}$ into two groups $A=\{\CNOT_{1,4}\}$, $B=\{\CNOT_{5,2}\}$, then for $g = \CNOT_{3,6}$, since $D(g, A) = 10 > D(g, B) = 6$, we add $\CNOT_{3,6}$ to group $A$. Finally, we get $A=\{\CNOT_{1,4}, \CNOT_{3,6}\}$, $B=\{\CNOT_{5,2}\}$. After obtaining the final groups, we schedule the group that has more gates (Line~\ref{line:more}) and leave the other to schedule later.

The following theorem ensures that this strategy is effective.

\begin{theorem}
    For a set of simultaneous two-qubit gates, perform \textsc{TwoQSchedule} multiple times until all the gates are scheduled. If they are scheduled into $K$ layers, then the top-$K$ closest gates in this set must belong to different layers.
\end{theorem}

\begin{myproof}
    Since scheduling once separates two closest gates, which will be executed in different layers, scheduling $K$ times leaves top-$K$ closest gates in different layers.\qed
\end{myproof}

Thus, when suppression performance is measured by distance, a decrease in parallelism necessarily leads to an improvement in suppression performance. This is because, for every additional layer, two more closest gates are guaranteed to be executed at different times.


\section{Evaluation}
\label{sec:evaluation}

We implement our approach as a general framework to work with different pulse optimization methods. In this section, we first give three methods that can achieve the $ZZ$-suppressing objectives defined in Section~\ref{sec:gate_lvl}, and then present their suppression performance. Next, we integrate them into our framework and evaluate our approach with QC benchmarks. Finally, we present the results of the Ramsey experiments performed on a real quantum computer.

\subsection{Methodology}

\subsubsection{Pulse Optimization Methods} \

\myleading{1. OptCtrl} With average gate fidelity $F_g$~\cite{nielsen2002simple} as the measure of similarity, the $ZZ$-suppressing objective~\ref{equ:obj-single} can be expressed as minimizing a loss function:
\begin{equation*}
    L = - F_g\left(U(T), U_1 \otimes \mathbf{I}^{\mathrm{Nbr}(a)}\right) - w F_g\left(U_{Ctrl}(T), U_1\right)
\end{equation*}
where the first $F_g$ term is to suppress cross-region $ZZ$ crosstalk, the second is to implement $U_1$, and $w$ is a weighting factor. The objective~\ref{equ:obj-two} can be converted similarly.

This is the typical setting of quantum optimal control~\cite{d2007introduction,ball2021software}. To suppress a range of $ZZ$ crosstalk strengths, we average the loss function values obtained at many different strengths.

\myleading{2. Pert} Optimizing the fidelity $F_g$ cannot reflect the order of $ZZ$ strength up to which crosstalk is suppressed. For direct and effective suppression, we propose an approach based on quantum perturbative theory.

The evolution according to $H(t) = H_{Ctrl}(t) + \lambda H_{Xtalk}$ can be written as $U(t) = U_{Ctrl}(t)U_{Xtalk}(t)$, where $U_{Xtalk}(t)$ can be expanded perturbatively according to the strength $\lambda$:
\begin{equation*}
\begin{aligned}
    U_{Xtalk}(t) &= I + \lambda U_{Xtalk}^\nb{1}(t) + O(\lambda^2) \\
    U_{Xtalk}^\nb{1}(t) &= - \frac{i}{\hbar} \int_0^t U_{Ctrl}^\dagger(t') \cdot H_{Xtalk} \cdot U_{Ctrl}(t') dt'
\end{aligned}
\end{equation*}

Thus, when $U_{Xtalk}^\nb{1}(T) = \mathbf{0}$, $U(T)$ will differ from $U_{Ctrl}(T)$ by $O(\lambda^2)$, i.e., the effect of the first order of $ZZ$ crosstalk is cancelled out. We achieve this by minimizing a loss function
\begin{equation*}
    L = \left\| U_{Xtalk}^\nb{1}(T) \right\| - w F_g\left(U_{Ctrl}(T), U\right)
\end{equation*}
where $U$ is the target gate to be achieved.

Both the loss functions of OptCtrl and Pert can be solved with gradient-based methods numerically~\cite{figueiredo2021engineering} or analytically~\cite{leung2017speedup}. We use the numerical method and set $T=$ 20ns.

\myleading{3. DCG} Dynamic corrected gates (DCG)~\cite{dcg1} are a generalization of dynamic decoupling~\cite{viola1998dynamical,viola1999dynamical}. Different from the above two approaches that optimize pulses from scratch, DCG constructs a sequence of pulses with existing ones (e.g., 20ns Gaussian pulses). The disadvantage of DCG is the long duration of the whole sequence (e.g., 120ns for $R_x(\pi/2)$ and 40ns for $I$).

\subsubsection{Native Gates} We compile quantum circuits to the same native gates as IBMQ devices~\cite{ibmq2021backend}, $\{R_z(\theta),\allowbreak R_x(\pi/2),\allowbreak R_{zx}(\pi/2)\}$, where $R_z(\theta)$ is virtual $Z$ gate implemented by software, not by pulses~\cite{mckay2017efficient}, and $R_{zx}(\pi/2)$ is used to implement CNOT gate~\cite{chow2011simple}. To suppress more crosstalk, we also optimize an identity gate $I = R_x(2\pi)$.

\subsubsection{Implementation} We implement our approach in Python. Simulations are performed at the Hamiltonian level with QuTiP~\cite{johansson2012qutip}, and Ramsey experiments are conducted on a real QC device~\cite{zhou2021rapid} with three transmon qubits.

\subsection{ZZ Crosstalk Suppression Performance}

We show in this section that pulses optimized under our $ZZ$-suppressing objectives can effectively suppress cross-region $ZZ$ crosstalk, even with control noise and leakage errors of qubits. We use Gaussian pulses as a reference since (i) they are representative of practical systems~\cite{motzoi2009simple,corcoles2015demonstration,ibmq2021backend} and (ii) they are not optimized for $ZZ$ crosstalk.

\subsubsection{Single-Qubit Gates}

For a two-qubit system \sBC1-\sC2, with crosstalk 1-2 fully suppressed, the evolution of the whole system would be $U \otimes I$ when applying a gate $U$ to qubit 1. Therefore, the infidelity between $U \otimes I$ and the actual observed evolution characterizes suppression performance.

\begin{figure}[!h]
    \centering
    \includegraphics[width=0.49\columnwidth]{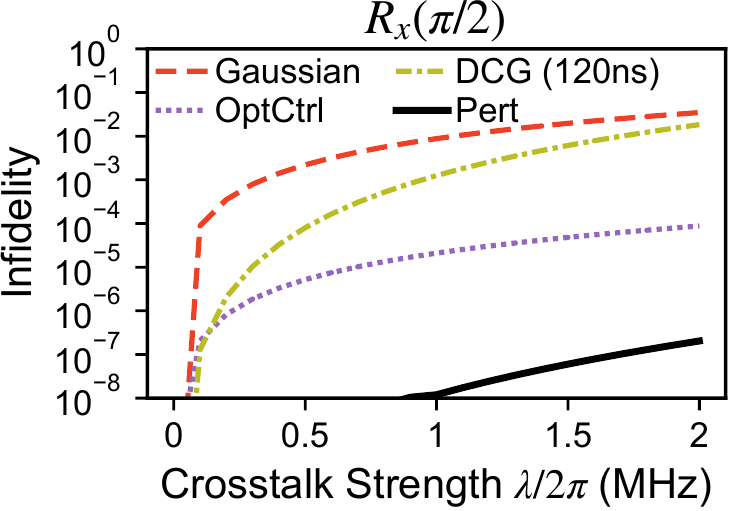}\ 
    \includegraphics[width=0.49\columnwidth]{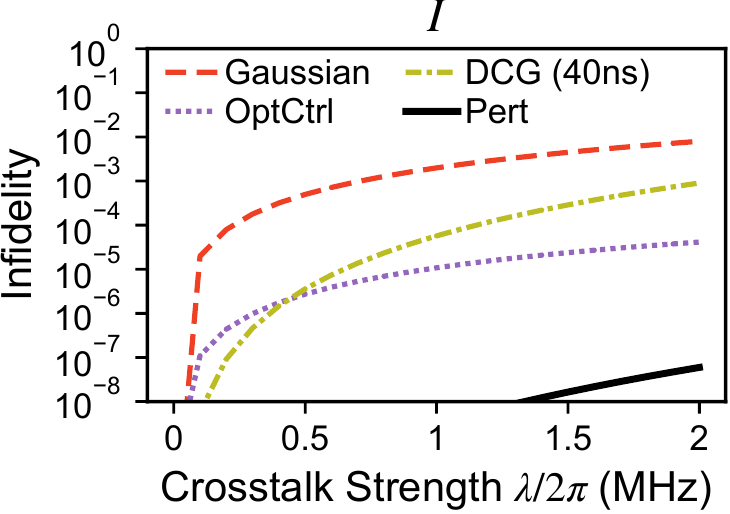}
    \caption{$ZZ$ crosstalk suppression performance of $R_x(\pi/2)$ and $I$ pulses. Lower is better. Precision is truncated to $10^{-8}$.}
    \label{fig:exp_single}
\end{figure}

\myparagraph{Suppression Performance} Figure~\ref{fig:exp_single} compares different pulses for native gates $R_x(\pi/2)$ and $I$. It is shown that all the optimized pulses achieve lower infidelity than Gaussian pulses. Since OptCtrl suppresses $ZZ$ indirectly via the observed fidelity, and the long DCG sequence accumulates more crosstalk errors than the others, they perform worse than Pert. However, as we will demonstrate later, for the typical $ZZ$ crosstalk strengths observed on real devices ($\lambda/2\pi \approx$ 200kHz~\cite{ash2020experimental,han2020error,sung2020realization,chow2011simple,andersen2020repeated}), OptCtrl is sufficient for our approach to achieve effective suppression of $ZZ$ crosstalk for the whole quantum circuit.

\begin{figure}[!t]
    \centering
    \begin{subfigure}[t]{0.47\columnwidth}
        \centering
        \includegraphics[width=\columnwidth]{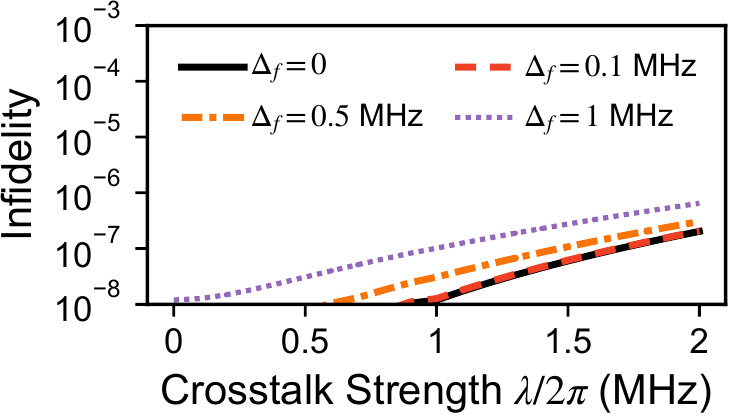}
        \captionsetup{justification=centering}
        \caption{Frequency detuning \\ ($\Delta_f = |f_{actual} - f_{desired}|$)}
    \end{subfigure}\quad
    \begin{subfigure}[t]{0.47\columnwidth}
        \centering
        \includegraphics[width=\columnwidth]{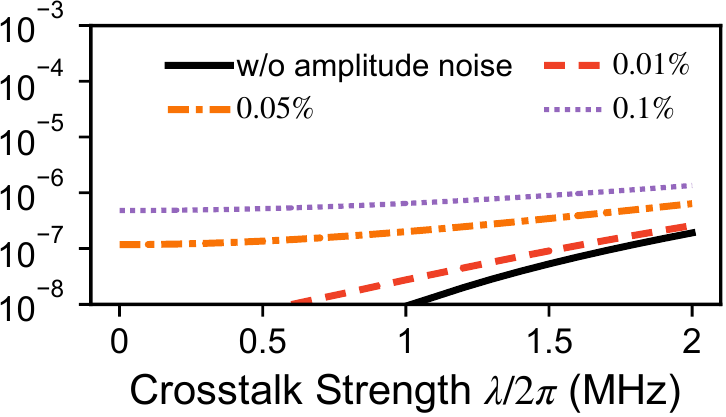}
        \caption{Amplitude Noise ($0.1\%$: amplitude fluctuates within $0.1\%$)}
    \end{subfigure}
    \caption{Robustness of the Pert $R_x(\pi/2)$ pulse to drive noise.}
    \label{fig:exp_detuning}
\end{figure}

\myparagraph{Drive Noise} Frequency detuning (shift) of carrier waves and amplitude fluctuation are two typical kinds of drive noise~\cite{leung2017speedup,carvalho2020errorrobust}. We show the results of the Pert $R_x(\pi/2)$ pulse in Figure~\ref{fig:exp_detuning}. It is shown that, under drive noise of typical strengths in practice (detuning < $0.1$ MHz, amplitude fluctuation < $0.1\%$~\cite{ball2021software}), $ZZ$ can still be effectively suppressed.

\begin{figure}[!t]
    \centering
    \includegraphics[width=0.95\columnwidth]{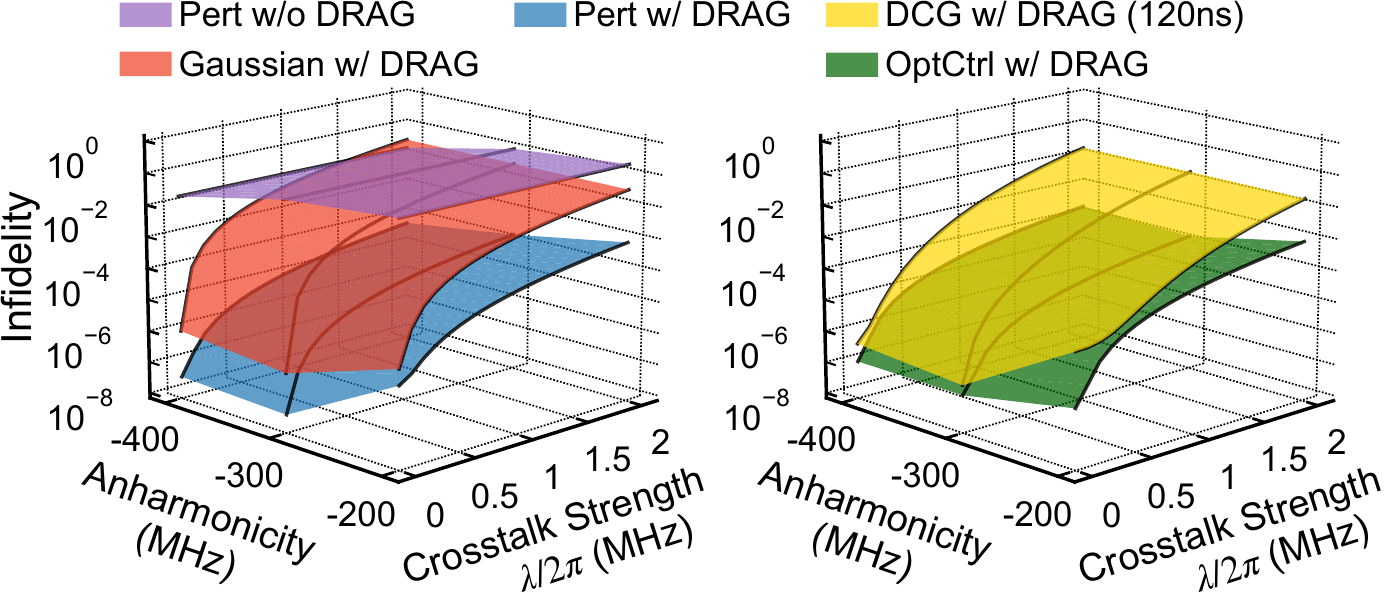}
    \caption{Suppression performance of $R_x(\pi/2)$ pulses under $ZZ$ crosstalk and leakage errors. Lower is better.}
    \label{fig:exp_leakage}
\end{figure}

Hamiltonian you optimize is a two-level system. DRAG is applied on top of that. 

\myparagraph{Leakage Errors} Superconducting qubits can suffer from leakage errors due to the existence of higher energy levels, an issue that can be mitigated using DRAG~\cite{gambetta2011analytic}. DRAG is a protocol that can modify pulses optimized for two-level systems and make them robust to leakage errors on multi-level systems. We process the optimized pulses with DRAG and then evaluate them on a five-level system with typical anharmonicity~\cite{arute2019quantum,chow2011simple}. Figure~\ref{fig:exp_leakage} shows that the optimized pulses processed by DRAG can achieve simultaneous suppression of both $ZZ$ crosstalk (versus Gaussian w/ DRAG) and leakage errors (versus Pert w/o DRAG). 

Since the effect of drive noise and leakage errors on the suppression performance is limited, we neglect these sources of errors and focus only on $ZZ$ crosstalk when performing the evaluation on quantum computing benchmarks.

\subsubsection{Two-Qubit Gates}

For a four-qubit system \sC1-\sBC2-\sBC3-\sC4 with a two-qubit gate on qubits 2 and 3, when (cross-region) crosstalk 1-2 and 3-4 are fully suppressed, the evolution of qubits 1 and 4 would be $I \otimes I$. We thus use the infidelity between $I \otimes I$ and the actual evolution to characterize suppression performance.

\begin{figure}[!h]
    \centering
    \begin{minipage}{1.04\columnwidth}
    \begin{subfigure}[t]{0.47\columnwidth}
        \centering
        \includegraphics[width=\columnwidth]{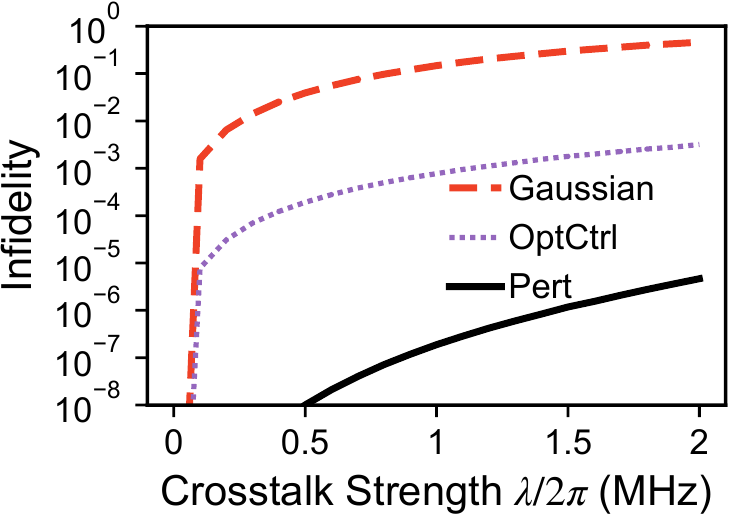}
        \caption{The same crosstalk strength on 1-2 and 3-4}
    \end{subfigure}\quad
    \begin{subfigure}[t]{0.47\columnwidth}
        \centering
        \includegraphics[width=\columnwidth]{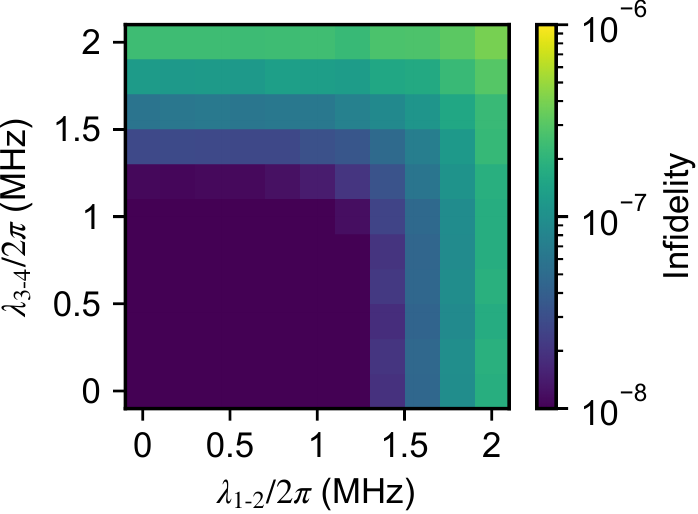}
        \caption{Different strengths on 1-2 and 3-4 (showing the Pert pulse)}
    \end{subfigure}
    \end{minipage}
    \caption{$ZZ$ crosstalk suppression performance of $R_{zx}(\pi/2)$ pulses. Lower is better.}
    \label{fig:exp_two}
\end{figure}

\myparagraph{Suppression Performance} Figure~\ref{fig:exp_two} shows the results under two configurations. (DCG is omitted since its sequence for two-qubit gates is too complicated and too long in practice.) These results show again the effective suppression of cross-region $ZZ$ crosstalk during two-qubit gates.

\subsection{Quantum Computing}
\label{sec:qc}

\begin{figure*}[!t]
\centering

\begin{minipage}{\columnwidth}
    \centering
    \includegraphics[width=\columnwidth]{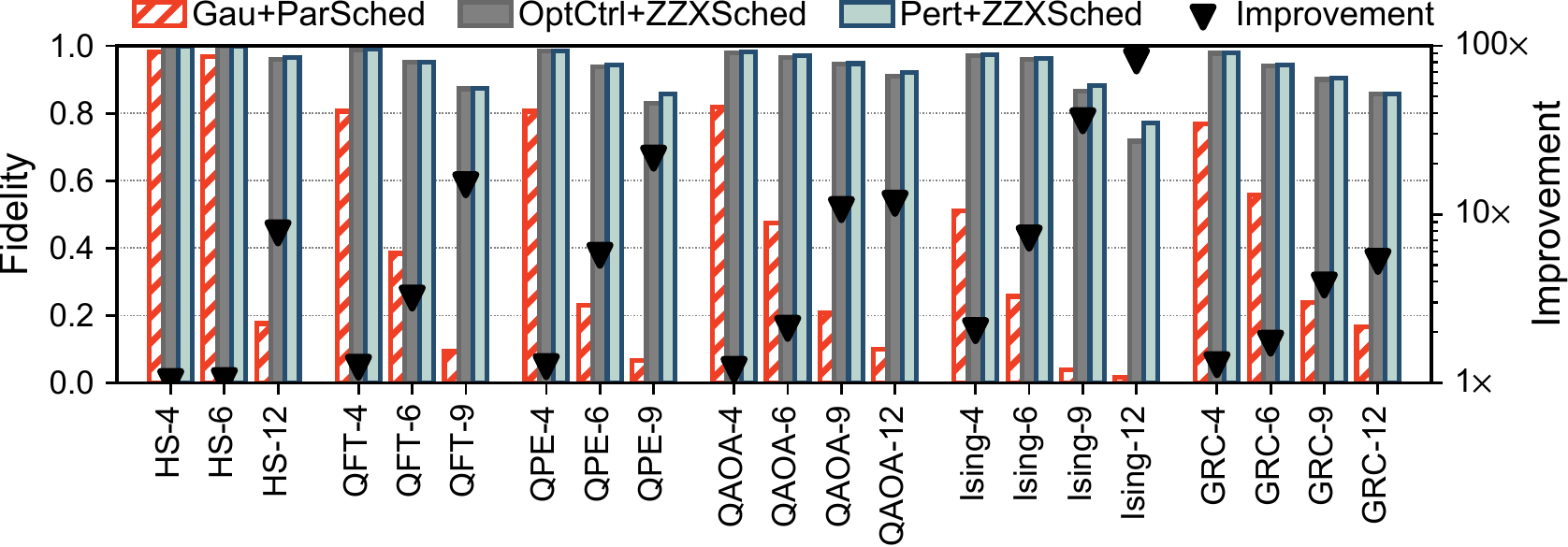}
    \caption{Overall improvements in fidelity under $ZZ$ crosstalk. Higher is better. (Gau: Gaussian pulses)}
    \label{fig:exp_app}
\end{minipage}\hfill
\begin{minipage}{\columnwidth}
    \centering
    \includegraphics[width=\columnwidth]{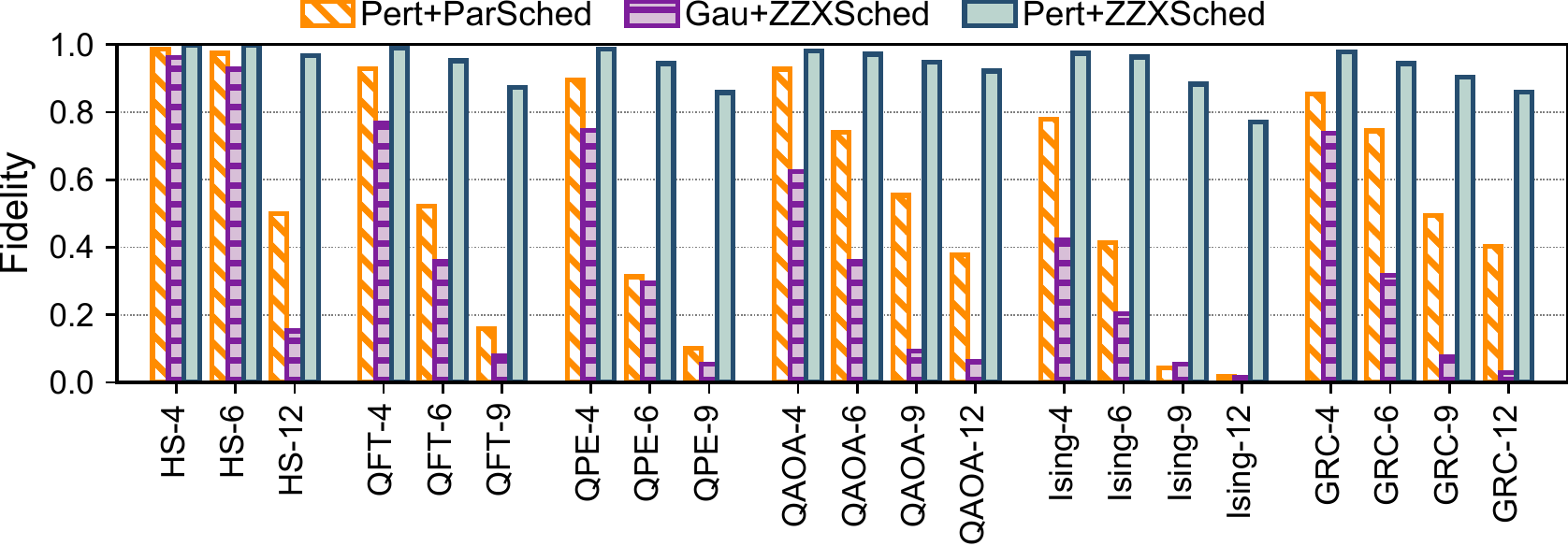}
    \caption{Comparisons with using only optimized pulses (Pert+ParSched) or ZZXSched (Gau+ZXSched).}
    \label{fig:exp_coopt}
\end{minipage}
\end{figure*}

\begin{figure*}[!t]
\centering

\begin{minipage}[t]{\columnwidth}
    \centering
    \includegraphics[width=\columnwidth]{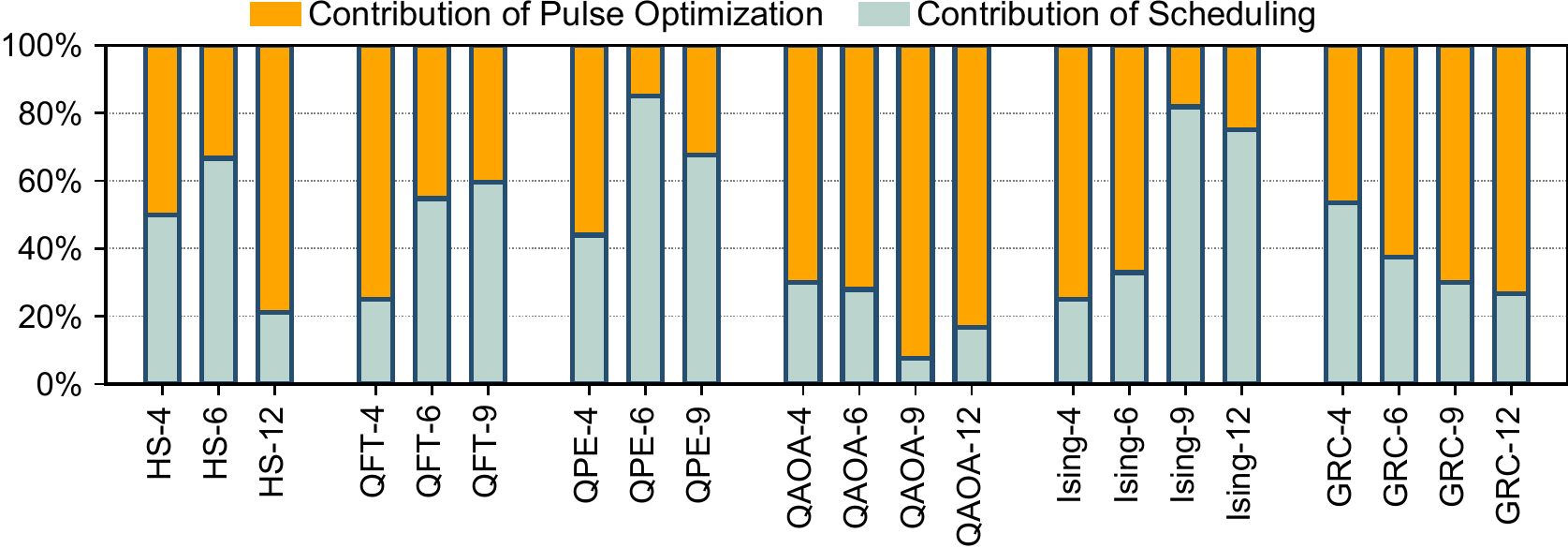}
    \caption{Contribution of pulse optimization/scheduling to the fidelity improvements from Gau+ParSched to Pert+ZZXSched.}
    \label{fig:exp_breakdown}
\end{minipage}\hfill
\begin{minipage}[t]{\columnwidth}
    \centering
    \includegraphics[width=\columnwidth]{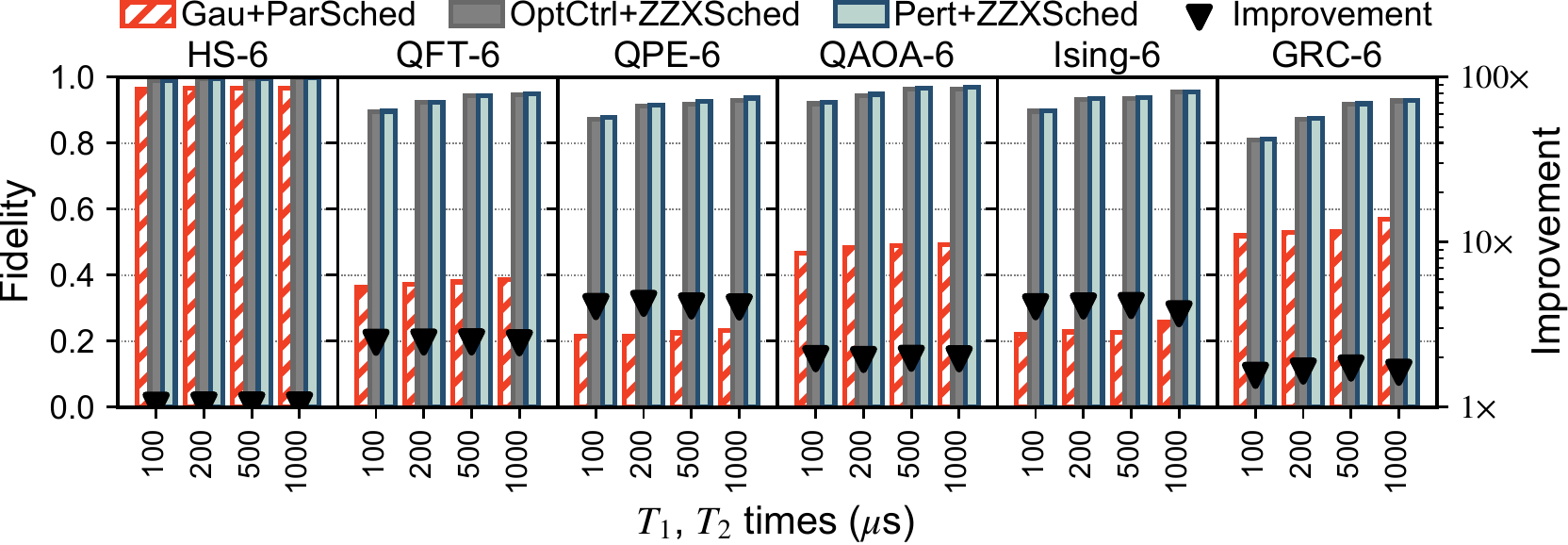}
    \caption{6-qubit benchmarks under $ZZ$ crosstalk and decoherence errors ($T_1 = T_2$). Higher is better.}
    \label{fig:exp_decoherence}
\end{minipage}

\end{figure*}

\myparagraph{Benchmarks} We use 6 representative benchmarks for near-term QC devices which have been widely used in recent works~\cite{murali2019noise,murali2019full,ding2020systematic,tannu2019not}: Hidden Shift (HS) algorithm~\cite{childs2005quantum}, Quantum Fourier Transform (QFT)~\cite{nielsen2002quantum}, Quantum Phase Estimation (QPE)~\cite{nielsen2002quantum}, Quantum Approximate Optimization Algorithm (QAOA)~\cite{farhi2014quantum}, Ising model simulation (Ising)~\cite{barends2016digitized}, and Google Random Circuits (GRC)~\cite{arute2019quantum}. We vary the number of qubits (4,6,9,12) for each benchmark. The compilation for any of these benchmarks with our approach takes <0.25s on a 2.3GHz CPU.

\myparagraph{Setup} Crosstalk strengths $\lambda/2\pi$ for different couplings are sampled from $N(\mu,\sigma^2)$ with $\mu$=200kHz and $\sigma$=50kHz~\cite{andersen2020repeated,ganzhorn2020benchmarking}. To analyze the effect of decoherence, we consider relaxation and dephasing characterized by the $T_1$ and $T_2$ times of qubits. We perform simulations on a $3 \times 4$ grid topology $G=(V,E)$, and the suppression requirement for scheduling is $N_Q < \max_{v \in V} degree(v)$ and $N_C \leq \frac{1}{2}|E|$. $\alpha=0.5$ and $k=3$ are used for the optimal suppression algorithm.

\myparagraph{Comparison} We name our scheduling policy ZZXSched and use it with OptCtrl and Pert pulses. We compare our approach with Gaussian pulses (Gau) + parallel scheduling (ParSched). ParSched maximizes the number of gates executed in parallel, which is the state of the art~\cite{murali2020software} in Qiskit~\cite{aleksandrowicz2019qiskit} and Qulic~\cite{smith2016practical}. We also make comparisons on devices with tunable couplers. The fidelity between actual output states and ideal output states (i.e., no error case) is used as the metric.

\myparagraph{Improvements in Fidelity} Figure~\ref{fig:exp_app} shows the overall improvements from pulse and scheduling co-optimization. There are three key results. \textbf{(1) Our approach can largely improve the circuit fidelity by up to 81$\times$} ($11\times$ on average) over Gau+ParSched. More importantly, we achieve > 0.9 fidelity for most of the evaluated benchmarks. \textbf{(2)} As revealed by the similar improvements with OptCtrl and Pert pulses, \textbf{our approach is insensitive to the pulses used}, as long as they can suppress $ZZ$ crosstalk to some extent. This is because, after compiled by our approach, circuits are affected mainly by the remaining intra-region crosstalk, and thus the varying performance of different pulses on cross-region crosstalk suppression becomes relatively unimportant. \textbf{(3) Our approach achieves an increasing improvement with more qubits.} This is significant, as it indicates that our approach can play a more important role in future larger devices.

\myparagraph{Analysis of Co-Optimization} Figure~\ref{fig:exp_coopt} shows the fidelity when using only optimized pulses (Pert+ParSched) or ZZXSched (Gau+\allowbreak ZZXSched). It highlights the synergistic effect of co-optimization by showing that co-optimization can achieve higher fidelity than using each part individually.

\myparagraph{Breakdown} Figure~\ref{fig:exp_breakdown} shows the contribution of pulse optimization and scheduling to the overall improvements. Averaged over all the tested benchmarks, pulse optimization and scheduling contribute 43.7\% and 56.3\% of the improvements respectively. The contribution of pulse optimization is calculated by the ratio of the improvement with only Pert pulses (Pert+ParSched) to the overall improvement.

\myparagraph{Analysis of Decoherence} In addition to $ZZ$ crosstalk, decoherence is also a significant source of errors. We therefore further evaluate our approach under both the two error sources. Figure~\ref{fig:exp_decoherence} presents the fidelity of 6-qubit benchmarks across a series of $T_1$ and $T_2$ times. It shows that our approach maintains stable improvements over different $T_1$ and $T_2$ times, revealing that the impact of decoherence on the effectiveness of our approach is limited.

\begin{figure*}[!t]
\centering

\begin{minipage}[t]{\columnwidth}
    \centering
    \includegraphics[width=\columnwidth]{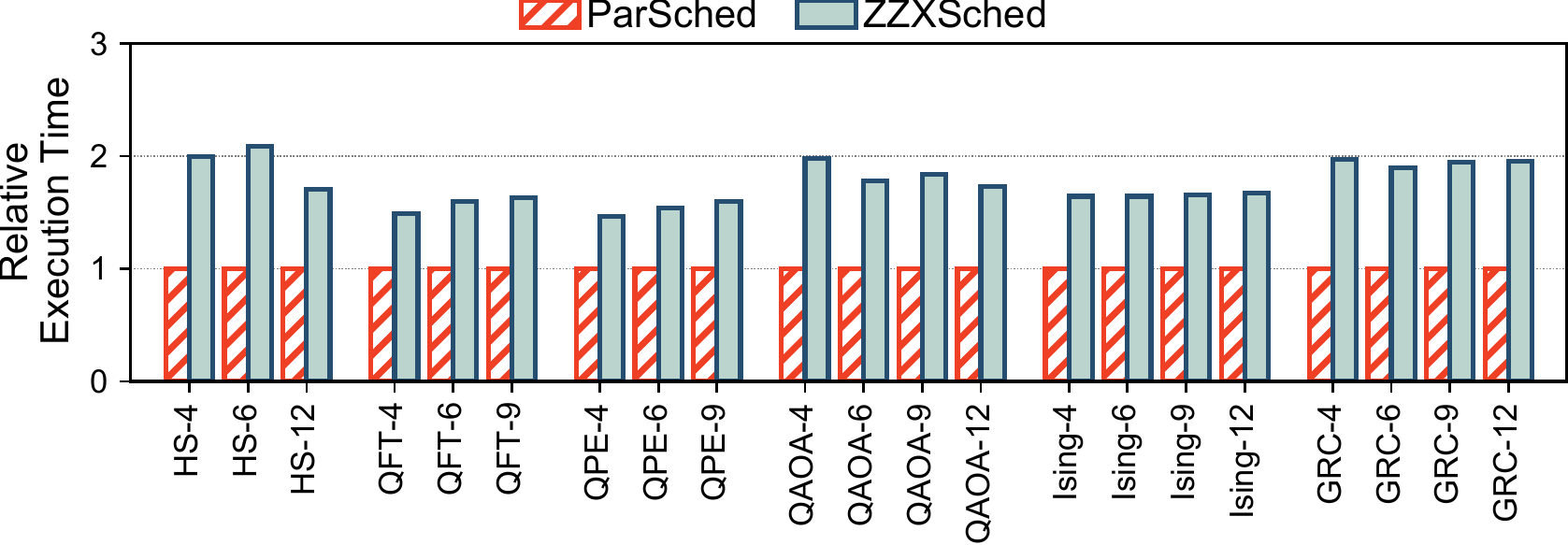}
    \caption{Execution time of benchmarks (relative to the time with ParSched). Results are irrelevant of pulses used.}
    \label{fig:exp_time}
\end{minipage}\hfill
\begin{minipage}[t]{\columnwidth}
    \includegraphics[width=\columnwidth]{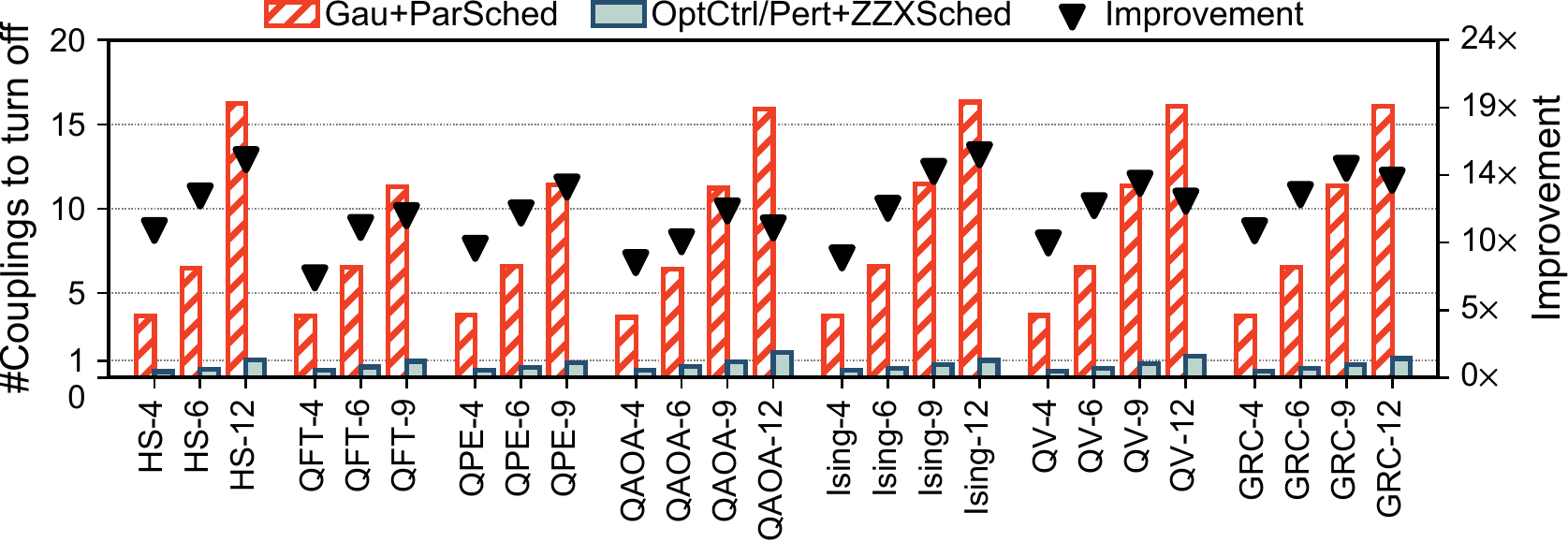}
    \caption{\#Couplings to turn off (averaged over all layers). Lower is better.}
    \label{fig:exp_turn_off}
\end{minipage}

\end{figure*}

\myparagraph{Effect on Parallelism} Figure~\ref{fig:exp_time} shows the execution time of each benchmark. Compared with ParSched, our ZZXSched typically increases the execution time by <2$\times$, showing a limited sacrifice of parallelism for better suppression. This trade-off is worthy, as we have already seen in both Figure~\ref{fig:exp_app} and \ref{fig:exp_decoherence} that the overall fidelity is improved.

\begin{figure*}[!t]
    \centering
    \includegraphics[width=0.9\textwidth]{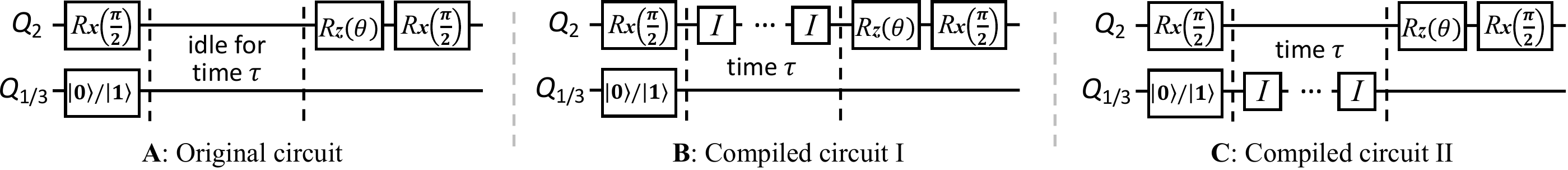}
    \caption{Circuits for Ramsey experiments on a real device $Q_1{-}Q_2{-}Q_3$. $\theta \propto \tau$. The difference between the compiled circuits B and C is that B applies $I$ gates to $Q_2$ while C applies $I$ to $Q_1$ and $Q_3$.}
    \label{fig:zz_exp_circuit}
\end{figure*}

\begin{figure*}[!t]
    \centering
    \includegraphics[width=0.9\textwidth]{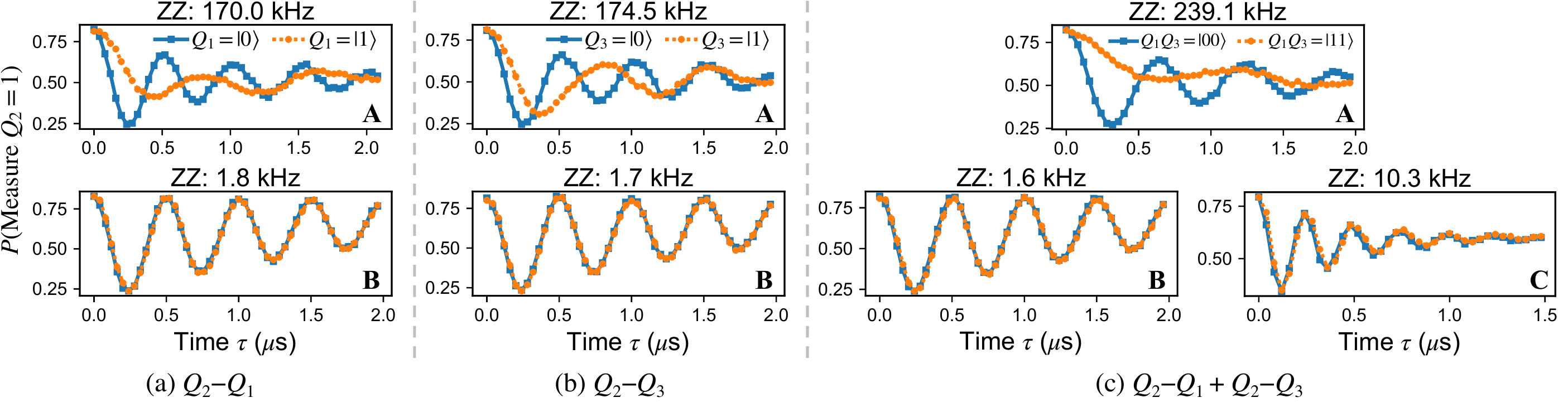}
    \caption{Results of Ramsey experiments. The letter at the lower right corner of each figure indicates the circuit from which the results are obtained. $ZZ$ strength is smaller when the oscillation frequency of the two curves gets closer.}
    \label{fig:zz_exp_result}
\end{figure*}

\myparagraph{Comparison on Devices with Tunable Couplers} Devices with tunable couplers can suppress $ZZ$ crosstalk by ``turning off'' couplings~\cite{mundada2019suppression}. In practice, the process of turning off can incur additional control noise~\cite{arute2019quantum,preskill2018quantum}. Our approach can be used on these devices and significantly reduce \#couplings to turn off (since only couplings within regions need to be turned off), thus reducing control noise and improving fidelity. Figure~\ref{fig:exp_turn_off} shows a $10{\sim}20\times$ reduction over the baseline and a very slow growth of \#couplings to turn off under our approach.

\subsection{Ramsey Experiments on a Real QC Device}
\label{sec:ramsey}

In this section, we further evaluate our approach from the perspective of effective $ZZ$ strength, where \textit{effective strength} refers to the strength that \textit{actually} affects the fidelity of quantum circuits. Thus, suppressing the effect of $ZZ$ crosstalk on QC is equivalent to reducing the effective $ZZ$ strength.

A standard protocol for measuring $ZZ$ strength between two qubits $Q_1, Q_2$ is to perform two Ramsey experiments on $Q_2$, with $Q_1$ in $\ket{0}$ or $\ket{1}$ respectively~\cite{chow2010quantum}. For each experiment, the population of $\ket{1}$ on $Q_2$ will oscillate at some frequency, and $ZZ$ strength can be calculated by the difference between the frequencies obtained from the two experiments. Figure~\ref{fig:zz_exp_circuit}~\textbf{A} shows the original circuit for Ramsey experiments. Our approach gives two compiled circuits that apply additional $I$ gates to different qubits, as shown in Figure~\ref{fig:zz_exp_circuit}~\textbf{B} and \textbf{C}.

We experiment on a real QC device~\cite{zhou2021rapid} with three transmon qubits in a line ($Q_1{-}Q_2{-}Q_3$). The qubit frequency is 5.78, 5.12 and 5.81 GHz respectively, and the adjacent qubits are coupled by a capacitor with the coupling strength ${\sim}$17 MHz. The default pulses on the device are Gaussian pulses, and our approach uses DCG pulses.

We have conducted three groups of experiments. Figure~\ref{fig:zz_exp_result} shows the experimental results. They reveal two important conclusions.

First, the results have verified the foundations of our approach. Figure~\ref{fig:zz_exp_result}~(a) and (b) show that cross-region crosstalk on one coupling can be suppressed, by pulses applied to one of the qubits on the coupling. Figure~\ref{fig:zz_exp_result}~(c) shows that cross-region crosstalk on multiple couplings around a qubit can be suppressed simultaneously, by pulses applied to either that qubit or its neighbors. These cases are the foundations of our approach, and the experiments have verified them.

Second, the results have shown the effectiveness of our approach. They show that our approach can reduce the effective $ZZ$ crosstalk strength from ${\sim}200$ kHz to <11 kHz, which indicates strong suppression of $ZZ$ crosstalk.


\section{Related Works}
\label{sec:related_work}

\myfirstparagraph{$ZZ$ Crosstalk Suppression} Previous approaches have mostly relied on sophisticated and specific chip design: tunable couplers~\cite{niskanen2007quantum,yan2018tunable,li2020tunable,sung2020realization}, heterogeneous qubits~\cite{ku2020suppression,zhao2020high,noguchi2020fast}, and multiple coupling paths~\cite{mundada2019suppression,kandala2020demonstration}. They can bring challenges to chip fabrication and introduce additional decoherence factors~\cite{kandala2020demonstration,malekakhlagh2020first}. We propose a \textit{software} approach that does not require special hardware to implement. It can be used on both the sophisticated chips described above (see Figure~\ref{fig:exp_turn_off}) and much simpler devices (see Figure~\ref{fig:zz_exp_result}). Our approach can potentially open up a new avenue to $ZZ$-resilient QC and simplify the design of QC devices.

\myparagraph{Error Mitigation} Pulse optimization~\cite{carvalho2020errorrobust,figueiredo2021engineering,gokhale2020optimized} or scheduling~\cite{murali2020software} has been explored by prior error mitigation works, but most of these works exploit only one of them. We take a step towards pulse and scheduling \textit{co-optimization} and show an effective application in $ZZ$ crosstalk suppression. We hope to inspire similar co-optimization strategies to tackle other types of errors.

Recently, two works have been proposed to mitigate other types of crosstalk than $ZZ$ crosstalk, XtalkSched~\cite{murali2020software} and ColorDynamic~\cite{ding2020systematic}. Our approach can be used with them to simultaneously mitigate multiple types of crosstalk. The following gives some potential ways to combine these approaches.

XtalkSched targets crosstalk arising when multiple gates are executed simultaneously. It appropriately inserts barriers in the circuit. Noting that these barriers actually divide a circuit into multiple sub-circuits, each sub-circuit can then be processed with our approach to further suppress $ZZ$ crosstalk.

ColorDynamic mitigates crosstalk due to frequency crowding. It annotates each gate with frequency information (for tuning qubit frequency dynamically) and groups gates into multiple layers. Similar to XtalkSched, each layer can be regarded as a circuit and then processed with our approach. 

\myparagraph{Dynamic Decoupling} Dynamic decoupling (DD) aims to protect a system from decoherence and dissipation due to unwanted system-environment interactions~\cite{viola1998dynamical,viola1999dynamical}. Though both DD and our approach target unwanted interactions, the main difference is that DD takes a view of idling, while we take a view of computing. Specifically, DD applies pulses during the \textit{idle} periods of qubits. These pulses are not aimed at implementing a target gate. In contrast, the primary goal of our pulses is to implement a target gate, though we also require the pulses to suppress $ZZ$ crosstalk surrounding the gate. With scheduling, we protect qubits in their whole lifetime including both computing and idle periods.

Our approach can use DD to provide protection during idle periods, by substituting DD pulses for the additional identity pulses. Actually, the DCG identity pulse is a special case of DD pulses; when using our approach with DCG pulses, the protection from DCG identity pulses can be regarded to be provided by DD.



\section{Conclusion}
\label{sec:conclusion}

We propose a pulse and scheduling co-optimization approach to suppress the destructive $ZZ$ crosstalk of superconducting QC devices in a scalable way. Our approach does not require special hardware to support. We have implemented it as a general framework and shown the compatibility with various pulse optimization methods. We have also demonstrated the effectiveness of our approach both in simulated settings and on a real device.

\section*{Acknowledgements}

We thank the anonymous reviewers for their valuable comments and suggestions. This work is under a joint project between Tsinghua University and Tencent Quantum Lab. This work is partially supported by National Key R\&D Program of China (2018YFA0306702) and by Key-Area Research and Development Program of Guangdong Province, under grant 2020B0303030002. 

\section*{A  APPENDIX: OPTIMIZED PULSES}
\label{sec:appendix}
\appendix

\begin{figure}[!ht]
    \centering
    \subfloat[OptCtrl\label{fig:optctrl}]{%
      \includegraphics[width=0.45\columnwidth]{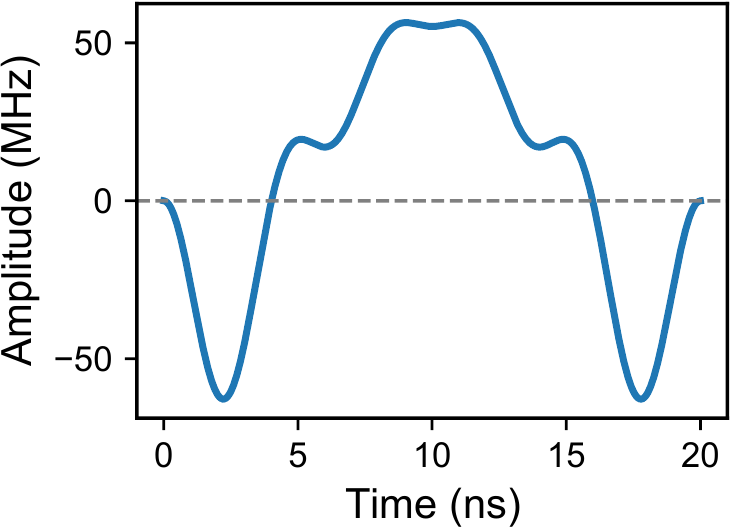}
    }
    \subfloat[Pert\label{fig:pert}]{%
      \includegraphics[width=0.45\columnwidth]{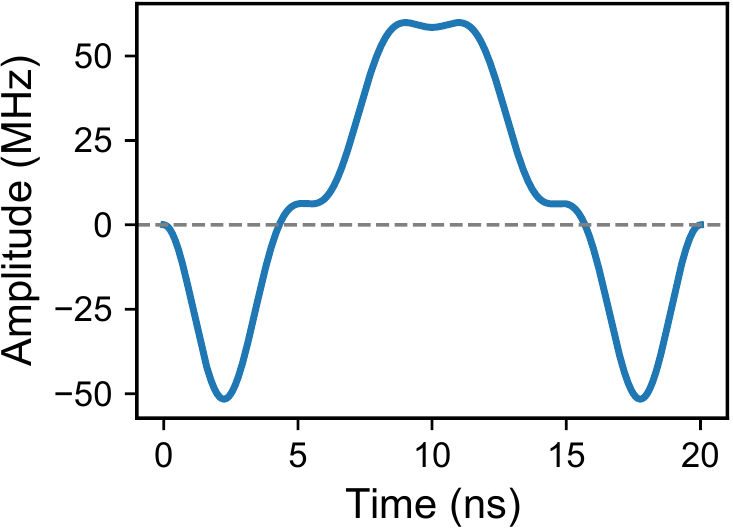}
    }
    
    \subfloat[DCG\label{fig:dcg}]{%
      \includegraphics[width=0.9\columnwidth]{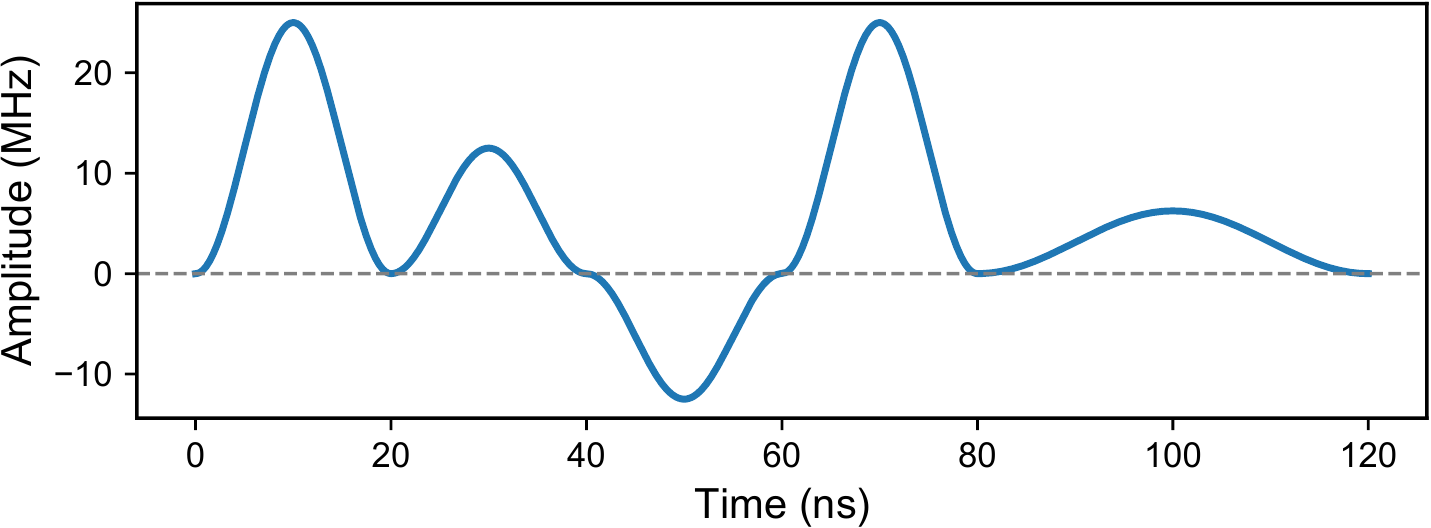}
    }
    \caption{Optimized pulses for $R_x(\pi/2)$. The amplitude and duration of these optimized pulses are reasonable~\cite{ball2021software}.}
    \label{fig:optimized_pulse}
\end{figure}

Figure~\ref{fig:optimized_pulse} shows the optimized pulses for $R_x(\pi/2)$.

We select a Fourier form $\Omega(\mathbf{A} ,t)$ for OptCtrl and Pert pulses, which is smooth, of narrow bandwidth and friendly to arbitrary waveform generators.
\begin{align*}
    \Omega(\mathbf{A} ,t) = \sum_{j=1}^{5} \frac{\mathbf{A}_j}{2} \left[1 + \cos\left(\frac{2\pi j}{T}t - \pi\right)\right]
\end{align*}
where $T$ denotes pulse duration, and $\mathbf{A} = (\mathbf{A}_1, \mathbf{A}_2, \cdots, \mathbf{A}_N)$ denote parameters to optimize.

DCG does not optimize pulses from scratch; it leverages existing pulses and combines them into a sequence~\cite{dcg1,dcg2}. Figure~\ref{fig:dcg} shows the sequence for $R_x(\pi/2)$ which is composed of 4 parts. (1) $0{\sim}20$ns: a Gaussian $\pi$ pulse; (2) $20{\sim}60$ns: a Gaussian $\pi/2$ pulse followed by a Gaussian $-\pi/2$ pulse; (3) $60{\sim}80$ns: a Gaussian $\pi$ pulse; (4) $80{\sim}120$ns: a Gaussian $\pi/2$ pulse.


\bibliographystyle{ACM-Reference-Format}


\end{document}